\documentclass[prb,twocolumn,showpacs,preprintnumbers,amsmath,amssymb,superscriptaddress]{revtex4-2}

\usepackage{graphicx}
\usepackage{mathrsfs}
\usepackage{dcolumn}
\usepackage{bm}
\usepackage{color}
\usepackage{bbm}
\usepackage[colorlinks=true,citecolor=blue,linkcolor=blue,urlcolor=blue]{hyperref}
\usepackage{svg}
\usepackage{soul}
\usepackage{rotating}
\usepackage{booktabs}
\usepackage{tablefootnote}
\begin{document}
\title{Chiral excitation flows of a multinode network based on synthetic gauge fields}
\author{Xian-Liang Lu}
\thanks{These authors contributed equally to this work.}
\affiliation{School of Physics, Sun Yat-sen University, Guangzhou 510275, China}
\affiliation{Department of Physics, University of Colorado, Boulder, CO 80309, USA}
\author{Fo-Hong Wang}
\thanks{These authors contributed equally to this work.}
\affiliation{School of Physics, Sun Yat-sen University, Guangzhou 510275, China}
\affiliation{School of Physics and Astronomy, Shanghai Jiao Tong University, Shanghai, 200240, China}
\author{Jia-Jin Zou}
\affiliation{School of Physics, Sun Yat-sen University, Guangzhou 510275, China}
\author{Ze-Liang Xiang}
\email{xiangzliang@mail.sysu.edu.cn}
\affiliation{School of Physics, Sun Yat-sen University, Guangzhou 510275, China}
\affiliation{State Key Laboratory of Optoelectronic Materials and Technologies, Sun Yat-sen University, Guangzhou 510275, China.}
% \date{\today}
%%%%
\begin{abstract}
Chiral excitation flows have attracted significant attention due to their unique unidirectionality. Such flows have been studied in three-node networks with synthetic gauge fields (SGFs), but the general theory of chiral flows in multinode networks requires further research and development. In this work, we propose a scheme to achieve chiral flows in $n$-node networks, where an auxiliary node is introduced to govern the system. This auxiliary node is coupled to all the network nodes, forming subtriangle structures with interference paths in these networks. We find the implicit chiral symmetry behind the perfect chiral flow and propose the universal criteria that incorporate previous models, facilitating the implementation of chiral transmission in various networks. By investigating the symmetries within these models, we present different features of chiral flows in bosonic and spin networks. Furthermore, we extend the four-node model into a ladder network, which is promising for remote state transfer in practical systems with reduced complexity. Our scheme can be realized in state-of-the-art experimental systems, such as superconducting circuits, magnetic photonic lattices, and ultracold atoms, thereby opening up possibilities for future quantum networks.
\end{abstract}
\maketitle
% ---------------------------------------------------------------------------
% ---------------------------------------------------------------------------
	
\section{Introduction}

Chiral excitation flow is a distinctive form of population transfer observed in quantum systems, that exhibits a unidirectional circulation characterized by chirality. A notable example of continuous chiral flows is dissipation-free edge currents in the quantum Hall effect~\cite{halperin1982quantized,Science2013Oh,NatRevPhys2020vonKlitzing}, which was first observed under a perpendicular magnetic field~\cite{Phys.Rev.Lett.1980Klitzing,Phys.Rev.B1981Laughlin} and later related to nontrivial topological effects~\cite{Shen2017TopologicalInsulatorsDirac}.
With the advancement of quantum technologies, edge currents can also be generated with synthetic gauge fields~\cite{dalibard2011colloquium,Goldman_2014,Jaksch_2003,Phys.Rev.A2010Koch,NaturePhoton2012Fang,PhysRevX.4.031031,Optica2015Schmidt,NewJ.Phys.2015Dubcek,Gorg2019,Phys.Rev.Lett.2021Chena,Nature2022delPino} in photonic systems~\cite{NaturePhoton2013Hafezi,Phys.Rev.X2016Anderson,Daniele2023}, ultracold quantum gases~\cite{atala2014observation,mancini2015observation,stuhl2015visualizing,tai2017microscopy,an2017direct,li2022atom} or other artificial systems~\cite{Phys.Rev.X2015Ningyuan,Phys.Rev.X2015Peano}. A discrete version of chiral excitation flows can also occur in artificial few-body systems characterized by broken time-reversal symmetry, parity symmetry, or both~\cite{Nat.Phys.2017Roushan,Nat.Phys.2019Wang,Appl.Phys.Lett.2020Liu}. In addition, they can serve as fundamental building blocks for creating larger artificial lattice models to study quantum Hall effects, such as the integer~\cite{NaturePhoton2013Hafezi} and fractional quantum Hall states~\cite{Nat.Phys.2017Roushan}.

Quantum networks connect various quantum systems via proven-secure channels and support quantum state transfer between separated nodes, which is potentially one of the most wide-ranging applications of current quantum technologies~\cite{Kimble2008,duan2010colloquium,reiserer2015cavity,cirac1997quantum,briegel1998quantum,Duan2001,fowler2010surface,Phys.Rev.Lett.2010Chudzicki,Phys.Rev.Lett.2011Pemberton-Ross,Ritter2012,Humphreys2018,Bhaskar2020,xiang2023universal}. In practice, the nodes designed for storing and processing quantum information in quantum networks can be realized with superconducting qubits~\cite{you2011atomic,devoret2013superconducting,Wendin_2017,GU20171,blais2021circuit}, atomic systems~\cite{Nature2008Greiner,duan2010colloquium,Rev.Mod.Phys.2010Saffman,Chien2015}, quantum dots~\cite{Rev.Mod.Phys.2007Hanson,Sci.Adv.2018Li}, or spin vacancies~\cite{PhysicsReports2013Doherty,Castelletto_2020,Phys.Rev.X2016Reiserer,Bradley2019}. Between these nodes, quantum channels play the role of bridges to coherently transfer information. In this framework, a chiral flow can provide site-by-site chiral transmission and enable unidirectional transfer of quantum states or entanglement in large-scale quantum networks. Chiral flows in few-node networks have been theoretically studied in various quantum systems, including superconducting circuits, photonic lattices~\cite{Phys.Rev.Lett.2021DeBernardis}, Rydberg-atom arrays~\cite{Phys.Rev.Research2022Wu}, giant atoms~\cite{du2023complex}, and cavity-magnonic systems~\cite{Phys.Rev.A2022Qi}, and they have even been experimentally achieved in both superconducting circuits~\cite{Nat.Phys.2017Roushan,Nat.Phys.2019Wang,Appl.Phys.Lett.2020Liu} and trapped-ion quantum processor\cite{shapira2023quantum}; however, these works demonstrated perfect chiral flows only in a triangular three-node network and fail to operate in larger networks. The systematic implementation of perfect chiral excitation flows in multinode networks—which could potentially lead to a wide range of applications in networks beyond a single triangular geometry—remains largely unexplored.

In this work, we develop a scheme to generate perfect chiral excitation flows within $n$-node networks based on synthetic gauge fields (SGFs), introducing an auxiliary node and proposing a mode for realizing the chiral dynamics. Moving beyond previous studies, we find the universal criteria for conducting chiral flows in different systems. For a three-node boson-SGF network, the broken time-reversal (TR) symmetry plays a critical role in the system's chiral dynamics, leading to complex hoppings between nodes, where the hopping phases cannot be simultaneously eliminated by a gauge transformation~\cite{Phys.Rev.A2010Koch}. By introducing an auxiliary node, we study the node-assisted SGF (aSGF) model with broken TR symmetry in each substructure of the system, in which the inherent relation between the $\pi/2$ hopping phase and a perfect chiral flow is unveiled. More generally, our aSGF model enables the emergence of these chiral flows within the $n$-node network by appropriately designing additional hoppings. In contrast to the bosonic system, when we generalize the SGF model to spin or two-level systems, where TR symmetry is preserved, chiral flows can still occur but exhibit opposite chiral dynamics for the up and down states. As a result, we study the fundamental elements for different models that support perfect chiral flows in the single or multiexcitation subspace and propose  universal criteria that incorporate previous models: (i) the symmetric and equally spaced energy spectrum in the corresponding excitation subspace and (ii) a complete set of chiral modes. In addition, we build an extended four-node network to function as a larger-scale network that also features chirality and can potentially be employed for directional remote transmission. To make these models feasible in practice, we investigate the influences of different types of imperfections in the aSGF network and introduce three experimental platforms for implementing our model: superconducting circuits, magnetic photonic lattices , and ultracold atoms. These platforms contribute to the chiral state transfer and    directional entanglement distribution for future quantum network and quantum simulation applications.

The remainder of theis paper is organized as follows. First, in Sec.~\ref{Sec:CEF}, we introduce the boson-SGF model and show the collapse of chiral excitation flow when generalizing previous studies to the four-node case. Thus, in Sec.~\ref{Sec:multinode}, we propose the aSGF model to rebuild the chiral flow in an $n$-node network by introducing an auxiliary node. We then investigate the spin-SGF model that preserves TR symmetry and supports opposite chiral dynamics and derive the general criteria for different models to conduct chiral flows. In Sec.~\ref{Sec:MNN}, we horizontally expand the four-node network to a large ladder, establishing a chiral flow between remote nodes with reduced complexity. In Sec.~\ref{Sec:IMP}, we discuss the sensitivity of the chiral flow to the imperfections in a four-node boson-SGF network with parameters taken from experimental data. The experimental feasibility of our scheme is considered in Sec.~\ref{Sec:Exper}. Finally, in Sec.~\ref{Sec:Con}, we summarize the main conclusions of this work and present the outlook for future applications.

% ---------------------------------------------------------------------------
% ---------------------------------------------------------------------------

\section{The chiral excitation flow}
\label{Sec:CEF}

\subsection{Synthetic gauge field model}

\begin{figure}[tbp]
\centering
\includegraphics[width=1.0\linewidth]{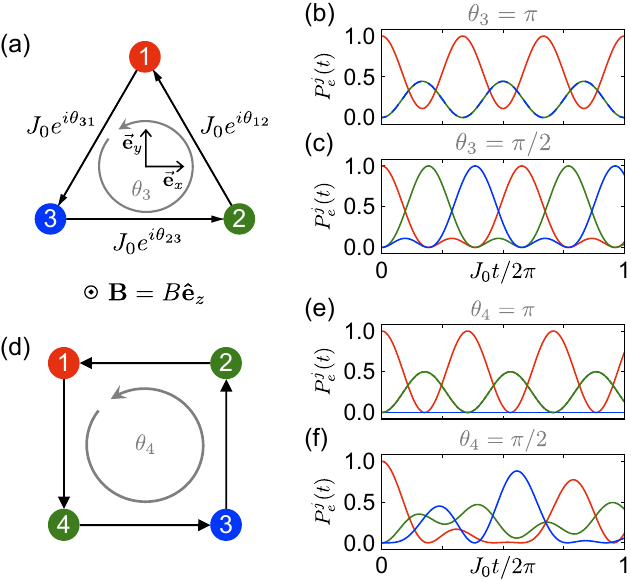}
\caption{(a) Illustration of the three-node boson-SGF model, where the synthetic flux $\theta_3$ determines the dynamical chirality. The evolution of excited-state populations $P_e^j(t)$ is shown for (b) $\theta_3=\pi$ and (c) $\theta_3=\pi/2$. The chiral excitation flow emerges when the synthetic flux of the loop $1\rightarrow2\rightarrow3$ is half-odd-integer multiples of $\pi$, yet the chirality disappears when the system preserves TR symmetry, i.e., $\theta_3=p\pi$ ($p\in \mathbb{Z}$). (d) Illustration of the four-node boson-SGF model. Similarly, (e) and (f) show the populations for $\theta_4=\pi$ and $\theta_4=\pi/2$, respectively. There is no chirality in the dynamics regardless of whether the synthetic flux $\theta_4$ breaks TR symmetry.}
\label{Fig1}
\end{figure}

To illustrate the concept of the chiral excitation flow, we first discuss the three-node bosonic network, as shown in Fig.~\ref{Fig1}(a). In this configuration, three boson modes with the identical frequency $\omega_0$ are positioned at the three nodes and coupled with their nearest-neighbor (NN) sites, forming a triangle arrangement in real space. Each mode can be described by the annihilation and creation operators $a_j$ and $a^{\dagger}_j$, where $j = 1,2,3$ labels each node site and the neighboring modes are coupled via the complex hoppings $J_{jk}=J_0e^{i\theta_{jk}}$. In the interaction picture, this three-node network is described by the following tight-binding Hamiltonian ($\hbar=1$)
\begin{equation}
\label{equ:SGFmodel}
H_{\rm SGF}^3=J_0\sum_{\langle jk\rangle}^3\left(e^{i\theta_{jk}}a_j^{\dagger}a_k+\mathrm{H.c.}\right),
\end{equation}
where the hopping phases $\theta_{jk}$ are induced from the SGF~\cite{Nat.Phys.2017Roushan,Phys.Rev.Research2022Wu}. We refer to this model as a three-node boson-SGF network, which can be readily extended to the $n$-node case with nodes arranged in a regular polygon [e.g., see Figs.~\ref{Fig1}(d), \ref{FigB1}(a), and \ref{FigB1}(b) for $n=4, 5$, and $6$, respectively].

% ---------------------------------------------------------------------------
% ---------------------------------------------------------------------------

The hopping phase is analogous to the Peierls phase acquired by a charged particle moving in a magnetic field. For an $n$-node network, the model can be visualized as a scenario in which a uniform magnetic field vertically penetrates the plane. In this scenario, we place the network on the $xy$ plane with its center located at the coordinate origin and apply a uniform synthetic magnetic field $\boldsymbol{B}=B(0,0,1)$ along the $z$ axis, as shown in Fig.~\ref{Fig1}(a). Thus, the effective vector potential $\boldsymbol{A}_\mathrm{eff}$ in the $xy$ plane induces a gauge-dependent hopping phase $\theta_{jk}=\frac{1}{\hbar}\int_{\boldsymbol{r}_k}^{\boldsymbol{r}_j}\boldsymbol{A}_\mathrm{eff}\cdot d\boldsymbol{r}$ between two nodes, while the synthetic flux threaded through the $n$-node polygonal network $\theta_n=\sum_{\langle jk\rangle,j<k}\theta_{jk}$ is gauge independent. In the following context, we mainly adopt the symmetric gauge with $\theta_{j,j+1}=\theta_n/n$, which greatly simplifies our derivation since the system exhibits the translation symmetry in this gauge. If one adopts another gauge characterized by a different set of hopping phases, the Hamiltonian is always connected to the symmetric gauge case through a gauge transformation and the results remain unchanged. In this system, the total excitation number, described by $\hat{n}_\mathrm{tot}=\sum_j a^\dagger_j a_j$, is a conserved quantity because it commutes with the tight-binding Hamiltonian, i.e., $[H_{\rm SGF}^{n},\hat{n}_\mathrm{tot}]=0$. Hence, the dynamics of the $n$-node SGF model can be confined to a specific excitation subspace and readily addressed using Schr\"{o}dinger equations even for large networks. For the single-excitation subspace, the dimension of the reduced Hilbert space is equal to the number of network nodes $n$.
%In a general case with a total excitation number of $n_e$, the dimension is denoted as $n!/n_e!(n-n_e)!$.

\subsection{Chiral flow in the single-excitation subspace}

The three-node chiral excitation flow has been investigated in different models~\cite{Nat.Phys.2017Roushan,Nat.Phys.2019Wang,Appl.Phys.Lett.2020Liu,Phys.Rev.Lett.2021DeBernardis}. Here, we focus on the single-excitation subspace of the boson-SGF model, where the first node is initially prepared in its excited state $|1\rangle$ and others are in their ground state $|0\rangle$. The wave function of the system is written as $|\psi(t)\rangle = \sum_{j=1}^n C_e^j(t) a^\dagger_j|G\rangle$, with initial condition $C_e^j(0)=\delta_{j1}$, where $|G\rangle\equiv\otimes_{j=1}^n|0_j\rangle$ is the ground state of the system. From the Hamiltonian in Eq.~\eqref{equ:SGFmodel}, we numerically calculate and plot the time evolution of the excited-state populations of the three bosonic modes. As shown in Fig~\ref{Fig1}(b), when the synthetic flux through the three-node loop is an integer multiple of $\pi$, i.e., $\theta_3=p\pi$ ($p\in\mathbb{Z}$), the system has TR symmetry~\cite{Phys.Rev.A2010Koch}, and the excitation will flow in both directions (clockwise and anticlockwise) with equal probability; however, when TR symmetry is broken, the excitation flow tends to be chiral. In particular, Fig.~\ref{Fig1}(c) shows that when $\theta_3$ is a half-integer multiples of $\pi$, i.e., $\theta_3=(p+1/2)\pi$ with $p\in\mathbb{Z}$, a perfect chiral excitation flow emerges. In this case, the excited-state populations $P^j_e(t)=|C_e^j(t)|^2$ can be analytically solved~\cite{Phys.Rev.Lett.2021DeBernardis,PhysicaE2004Taut}: 
\begin{equation}\label{equ:3nodeSGF_popu}
P^j_e(t)=\left\{ \frac{1}{3}+\frac{2}{3}\cos\left[\sqrt{3}J_0t-\frac{2\pi}{3}\left(j-1\right)\right]\right\}^2.
\end{equation}
It takes the dimensionless time $J_0T_0=2\pi/3\sqrt{3}\approx 1.21$ to fully transfer the excited-state population from site $j$ to site $(j+1)~\text{mod}3$. Furthermore, the dynamics of the excited-state populations for two nearest-neighboring nodes, e.g., $P^1_e(t)$ and $P^2_e(t)$, turns out to have a $2\pi/3$ phase shift. This feature explicitly interprets Fig.~\ref{Fig1}(c) and implies the dynamical chirality of this network. We also note that as $\theta_3 = (p+1/2)\pi$ in which $p\in\mathbb{Z}$, the system exhibits a symmetric spectrum in the single-excitation subspace with energy levels $E=\{-\sqrt{3}J_0,0,\sqrt{3}J_0\}$, which reveals an underlying chiral symmetry, namely, $\mathcal{C}^{-1}\mathcal{H}\mathcal{C}=-\mathcal{H}$. This usage of the term ``chiral'' differs from its previous context in ``chiral flow'', where ``chiral'' refers to the lack of inversion symmetry in excitation flows. We will revisit this point frequently when explaining and designing multinode chiral flows in Sec.~\ref{Sec:multinode}.

However, after a simple generalization, the SGF model fails to achieve such chiral flows in networks consisting of four or more nodes~\cite{PhysicaE2004Taut,Nat.Phys.2019Wang}. In Figs.~\ref{Fig1}(e) and~\ref{Fig1}(f), we plot the time evolution of the populations for a four-node SGF network. When the synthetic flux through this four-node loop ($\theta_4=\pi/2$) breaks TR symmetry, the excitation dynamics still show no chirality. In fact, the populations of the second and fourth nodes are always the same, $P^2_e(t)=P^4_e(t)$, with details and analytical proof given in Appendix~\ref{asec:4nodeSGF_nocentral}. If the synthetic flux $\theta_4=\pi$, the population of the third node vanishes, and it can thus be referred to as a ``dark site''~\cite{Arimondo1996CoherentPopulationTrapping,Phys.Rev.Lett.2000Fleischhauer}. This results from the Aharonov-Bohm effect due to the destructive quantum interference from the paths $1\rightarrow 2\rightarrow 3$ and $1\rightarrow 4\rightarrow 3$. Further, as discussed in Appendix~\ref{asec:5nodeand6node_SGF}, theoretical analysis suggests that the general $n$-node ($n\geq 4$) SGF model cannot support a perfect chiral flow.

% ---------------------------------------------------------------------------
% ---------------------------------------------------------------------------

\section{Chiral flows in multinode networks}
\label{Sec:multinode}

\subsection{Perfect chiral flow assisted by an auxiliary node in a four-node network}
\label{Sec:FNN}

\begin{figure}[tbp]
\centering
\includegraphics[width=1.0\linewidth]{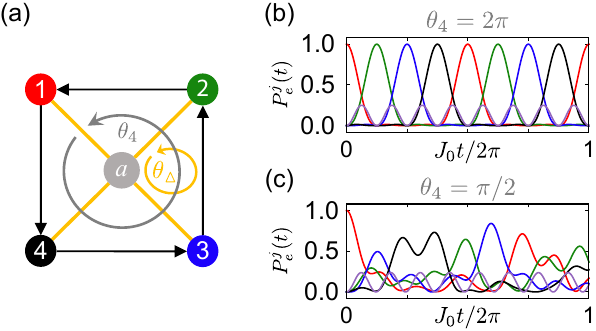}
\caption{(a) Four-node network assisted by an auxiliary node, referred to as the four-node boson-aSGF model. Each network node is connected to the auxiliary node with the coupling $\beta_c J_0$. (b) Evolution of the excited-state populations for $\theta_4=2\pi$. In each subtriangle structure, TR symmetry is broken with the synthetic flux $\theta_\vartriangle=\pi/2$. (c) Evolution of the excited-state populations for $\theta_4=\pi/2$. In contrast to the three-node SGF model, the dynamics show no chirality with the $\pi/2$ total flux.}
\label{Fig2}
\end{figure}

Motivated by the quantum interference of the two paths in the four-node SGF model, here, we introduce an auxiliary node that is positioned at the center of the network and couples to all network nodes, dividing the network into four subtriangles, as depicted in Fig.~\ref{Fig2}(a). In this case, with appropriate parameters, we demonstrate that the quantum interference assisted by subtriangles from different paths is able to induce a perfect chiral excitation flow.

The centrally located auxiliary node can be described by annihilation and creation operators $c$ and $c^\dagger$. For simplicity, this auxiliary node is assumed to homogeneously couple to other nodes with a real-valued hopping rate. The Hamiltonian of the node-assisted SGF (aSGF) model reads
\begin{equation}
\label{equ:SGFCmodel}
H_\mathrm{aSGF}^4/J_0=\sum_{\langle jk\rangle}^4e^{i\theta_{jk}}a_j^\dagger a_k +\beta_c\sum_{j=1}^4 a^\dagger_j c+\mathrm{H.c.},
\end{equation}
where $\beta_c$ characterizes the hopping strength between the auxiliary node and other network nodes. This aSGF Hamiltonian will return to $H_{\rm SGF}^4$ as $\beta_c=0$. Here, we focus on the single-excitation subspace, and the system's time-dependent wave function can be written as $|\psi(t)\rangle = \sum_{j=1}^N \left[C_e^j(t)a_j^\dagger+C_e^c(t)c^\dagger\right]|G\rangle$, where $|G\rangle\equiv(\otimes_{j=1}^N|0_j\rangle)\otimes |0_c\rangle$ is the new ground state, and we set the initial state as $C_e^j(0)=\delta_{j1}$ and $C_e^c(0)=0$. The dynamics of the system can be analytically solved, and a detailed derivation can be found in Appendix~\ref{asec:deri}. Remarkably, a perfect chiral flow occurs when the central coupling strength is given by $\beta_c=2$ and the synthetic flux through the four-node loop is $\theta_4=2\pi$. This optimal value can be directly found in the solution of excited-state populations $P_{e}^{j}=|C_e^j(t)|^2$, where
\begin{equation}
\label{equ:4nodespopu}
\begin{aligned}
C_{e}^{j}\left(t\right)=&\frac{1}{2}\cos\left[2J_{0}t+\frac{(j-1)\pi}{2}\right]\\
&\qquad\quad+\frac{1}{4}\cos\left(2\beta_c J_0 t\right)+\frac{(-1)^{j-1}}{4}.
\end{aligned}
\end{equation}
When $\beta_c=2$, the two cosine functions in Eq.~\eqref{equ:4nodespopu} can simultaneously reach their peak values, indicating the complete transmission of the excited-state population between neighboring nodes. In particular, this configuration of hopping phases ($\theta_{j,j+1}=\pi/2$, and the hopping phase to the auxiliary node is $0$) can still be interpreted as a vertical uniform magnetic field with a symmetric gauge chosen. Gauge invariance can be verified numerically: If one instead works in Landau gauge, $\boldsymbol{A}=(-By,0,0)$, the same dynamical results will be obtained. 
% ---------------------------------------------------------------------------
% ---------------------------------------------------------------------------

To intuitively understand the underlying cause of the chiral flow that emerges in this auxiliary-mediated network, it is instructive to focus on subtriangles inside the four-node pattern. The pure four-node boson-SGF model in Fig.~\ref{Fig1}(d) does not support a chiral flow due to the absence of triangular substructures; however, as shown in Fig.~\ref{Fig2}(a), the auxiliary node can provide four subtriangles that share the same hopping phase (or flux) $\theta_\vartriangle=\theta_4/4$. When we set $\theta_\vartriangle$ as the phase that completely breaks TR symmetry in the three-node boson-SGF model, e.g., $\theta_\vartriangle=\pi/2$, the perfect chiral flow between these four nodes indeed occurs, which implies that the three-node structure acts as a fundamental unit dominating the expected four-node chiral flow. In contrast, in Fig.~\ref{Fig2}(c), we show a completely different scenario in which the chiral flow collapses as $\theta_4=\pi/2$. This demonstrates that achieving a perfect chiral flow requires broken TR symmetry in each subtriangle rather than in the whole network. In fact, the hopping phase $\theta_{j,j+1}=\pi/2$ ensures a chiral symmetry that facilitates the chiral flow, as discussed in the next subsection.

Further, the auxiliary node also contributes to the formation of chiral flow by mediating the energy spectrum and removing degeneracy. The energy levels of the four-node boson-SGF model are given by $E_{1,\pm}=\pm 2J_0$ and $E_{0,\pm}=0$, which results in imperfect oscillations at even sites (see Appendix~\ref{asec:5nodeand6node_SGF} for details), as shown in Fig.~\ref{Fig1}(e). Due to the internal chiral symmetry of $H^4_{\rm SGF}$, the degenerate zero-energy eigenstates are in pairs, given by $|\varphi_{0,+}\rangle=(1,0,1,0)^T/\sqrt{2}$ and $|\varphi_{0,-}\rangle=(0,1,0,1)^T/\sqrt{2}$; 
however, in the presence of the auxiliary node, these two states are lifted, along with the emergence of a new zero-energy state. The chiral symmetry also persists as the auxiliary node is introduced, with the single-particle form of the chiral symmetry operator $\mathcal{C}$ written as
\begin{equation}
\mathcal{C}^{-1}\mathcal{H}_\mathrm{aSGF}^{4}\mathcal{C}=-\mathcal{H}_\mathrm{aSGF}^{4}, \quad \mathcal{C}=
\begin{pmatrix}
1&0&0&0&0\\0&0&0&1&0\\0&0&1&0&0\\0&1&0&0&0\\0&0&0&0&-1
\end{pmatrix},
\end{equation}
where $\mathcal{H}_\mathrm{aSGF}^4$ represents the single-particle Hamiltonian [see Eq.~\eqref{equ:n_node_Hcal_aSGF} with $\theta_{j,j+1}=\pi/2$ for the matrix form]:
\begin{equation}
H_\mathrm{aSGF}^4 = S^\dagger \mathcal{H}_\mathrm{aSGF}^4 S,\quad S^\dagger=(a_1^\dagger,a_2^\dagger,\cdots, c^\dagger).
\end{equation}
This configuration of hopping phases ensures the chiral symmetry of the aSGF model. The chiral transformation is equivalent to inverting the network about the diagonal “1a3” [see Fig.~\ref{Fig2}(a)] and performing the transformation $c^\dagger\rightarrow -c^\dagger$ at the same time. Under this transformation, all the hoppings reverse their directions, leading to an overall minus sign to the single-particle Hamiltonian.
Thus, the nonzero energy eigenstates emerge in pairs $\{|\varphi\rangle,\mathcal{C}|\varphi\rangle\}$ with opposite eigenenergies, while the sole unpaired state should be a zero-energy eigenstate~\cite{Asboth2016ShortCourseTopological}. In Eq.~\eqref{equ:4nodespopu}, the lifted energies $E_{\beta_c,\pm}=\pm 2\beta_c J_0$ result in an extra standing wave to this network, and the zero-energy state is denoted by $|\varphi_0^\prime\rangle=\sum_{j=1}^4 \frac{1}{2\sqrt{2}}a_j^\dagger|G\rangle+\frac{1}{\sqrt{2}}c^\dagger |G\rangle$, contributing the constant $(-1)^{j-1}/4$ as described in the last term. To ascertain the value of $\beta_c$ and validate the feasibility of conducting the chiral flow from a physical perspective, we rewrite the probability amplitudes in Eq.~\eqref{equ:4nodespopu} as 
\begin{equation}
\label{4noderewrite}
\begin{aligned}
C_e^j(t)=&(-1)^{(j-1)}\left\{\frac{1}{2}\cos\left[2J_{0}t-\frac{(j-1)\pi}{2}\right]\right.\\
&\left.+\frac{1}{4}\cos\left[2\beta_cJ_0t-(j-1)\pi\right]+\frac{1}{4}\right\}.
\end{aligned}
\end{equation}
In this sense, the dynamical superposition from lifted eigenstates $|\varphi_{\beta_c,\pm}\rangle$ is equivalent to a $\pi$-shift traveling wave with frequency $2\beta_c 
J_0$, and the zero-energy state $|\varphi_0^\prime\rangle$ supplies the last component for a perfect oscillation. From Eq.~\eqref{4noderewrite} we verify that the perfect chiral flow is achieved if and only if $\beta_c=2$.

In short, we conclude that the subtriangle structure's TR symmetry determines the dynamical chirality of the bosonic multinode network. When the additional auxiliary node is introduced to couple to network nodes, and the synthetic flux of subtriangles $\theta_\triangle$ is chosen to be the phase referred to in the three-node case, i.e., $(p+1/2)\pi$ with $p\in\mathbb{Z}$, the perfect chiral flow can emerge with a specific value of $\beta_c$. In particular, the $\pi/2$ hopping phase and the auxiliary node are the two essential elements for achieving the perfect chiral flow in the four-node boson-aSGF model; however, when we directly extend the boson-aSGF model to $n$-node networks, due to the anharmonicity of energy spectrum, the chiral flow collapses regardless of the strength of $\beta_c$ (as detailed in Appendix~\ref{asec:5nodeand6node_SGF}), and explicit expressions for a six-node aSGF model can be found in Appendix~\ref{asec:deri}. Nevertheless, the study of the four-node network demonstrates the principle of conducting chiral flows from the quantum interference of different states. In the next subsection, we will show that by introducing the next-nearest-neighbor (NNN) hopping in five- and six-node networks, perfect chiral flows can be revived with the boson-aSGF model.

\subsection{Perfect \texorpdfstring{$n$}{}-node chiral flow beyond NN interactions}\label{n_node_chiral}

Since the pure $n$-node boson-SGF model does not support the chiral flow, we investigate the prerequisites necessary for achieving a perfect chiral flow in the $n$-node boson-aSGF network. When an auxiliary node is coupled to network nodes, it only affects the zero-energy state(s). In the content that follows, we use periodic boundary conditions with $a_{n+j}=a_{j}$. The system Hamiltonian is now described by $H_\mathrm{aSGF}^{n} / J_0=\sum_{j=1}^n(i a_{j}^\dagger a_{j+1}+\beta_c a_j^\dagger c)+\text {H.c.}$,
which can be block-diagonalized using the eigenstates of the $n$-node boson-SGF Hamiltonian (see Appendix~\ref{asec:5nodeand6node_SGF} and details of four- and six-node cases in Appendix~\ref{asec:deri}). Analogous to the result for the $n$-node boson-SGF model, the $\pi/2$ hopping phase yields a chiral symmetry for the single-particle Hamiltonian $\mathcal{H}_\mathrm{aSGF}^{n}$:
\begin{equation}
\mathcal{C}^{-1} \mathcal{H}_\mathrm{aSGF}^{n} \mathcal{C}=-\mathcal{H}_\mathrm{aSGF}^{n}, \quad 
\mathcal{C}=\begin{pmatrix}
1&0&\cdots&0&0\\
0&0&\cdots&1&0\\
\vdots&\vdots& \reflectbox{$\ddots$}&\vdots&\vdots\\
0&1&\cdots&0&0\\
0&0&\cdots&0&-1
\end{pmatrix},
\end{equation}
leading to a symmetric energy spectrum for the system with paired eigenstates. Due to the disparity in zero-energy eigenstates, we separately analyze the scenarios for even and odd values of $n$. With identical couplings between the auxiliary node and $n$ network nodes, the Hamiltonian can be written as
\begin{equation}\label{equ:intermediate_diag}
\mathcal{H}_\mathrm{aSGF}^n=\sum_{\substack{m\neq 0}}E_m |\varphi_m\rangle \langle \varphi_m| + \sqrt{n}\beta_c J_0 |\varphi_0\rangle\langle c| + \text{H.c.},
\end{equation}
where $|\varphi_m\rangle~(-n/2<m\leq n/2)$ are the eigenstates of $\mathcal{H}_\mathrm{SGF}^n$, as discussed in Appendix~\ref{asec:5nodeand6node_SGF}, and $|c\rangle=c^\dagger |G\rangle$ is the excited state of the auxiliary node. Here, the energy spectrum $E_m=-2J_0\sin k_m$ exhibits the dispersion relation of the one-dimensional tight-binding Hamiltonian, where $k_m=2\pi m/n$ are wave numbers in the first Brillouin zone. We refer to these eigenstates $|\varphi_m\rangle$ as chiral modes in this system because they have chiral phase shifts $2\pi m/n$ between nearest sites. For an odd value of $n$, the zero-energy state $|\varphi_{0}\rangle$ hybridizes with the product state $|c\rangle$, and they eventually form the new paired chiral modes $|\varphi_{\beta_c,\pm}\rangle=(|\varphi_0\rangle \pm |c\rangle)/\sqrt{2}$, which are independent of the value of $\beta_c$, while the eigenenergies $\pm E_{\beta_c}=\pm\sqrt{n}\beta_c J_0$ are tuned by $\beta_c$. If $n$ is an even number, the chiral symmetry of $H_\mathrm{SGF}^n$ ensures the existence of two degenerate zero-energy states, corresponding to $|\varphi_{0}\rangle$ and $|\varphi_{n/2}\rangle$. Similarly, the auxiliary node results in hybridization between the two states $|\varphi_{0}\rangle$ and $|c\rangle$, and the degeneracy is thus lifted. Instead, $|\varphi_{n/2}\rangle$ remains as the unperturbed zero-energy state. We emphasize that the hybridized states, together with those unperturbed states $|\varphi_{m}\rangle$, form a complete set of required chiral modes, which makes it possible to realize a perfect chiral flow in $n$-node networks.
% ---------------------------------------------------------------------------
\begin{figure}[tbp]
\centering
\includegraphics[width=1.0\linewidth]{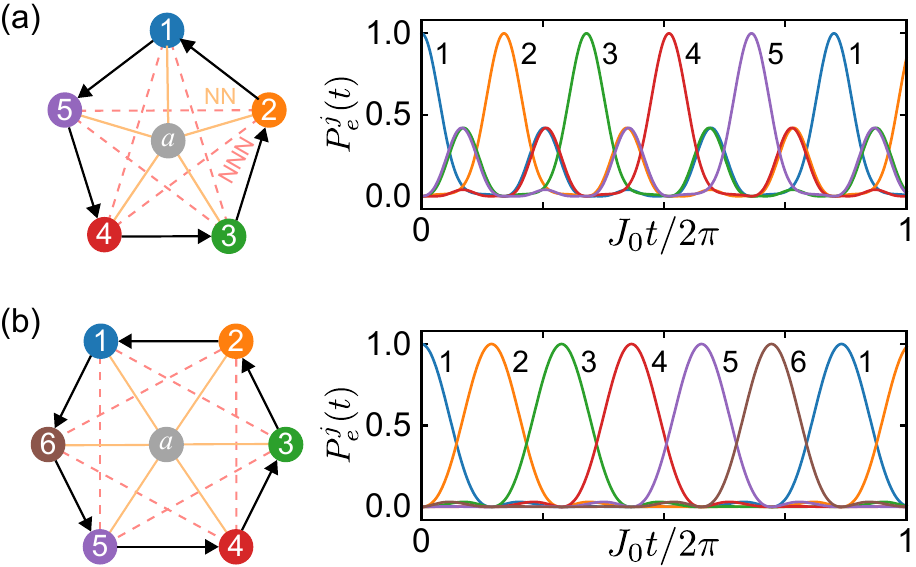}
\caption{Dynamics of the modified (a) five-node and (b) six-node aSGF models. NNN hopping is introduced in the modified networks to ensure perfect chiral flows.}
\label{Fig3}
\end{figure}
% ---------------------------------------------------------------------------

We now deduce the Hamiltonian from the prerequisites to create a perfect chiral flow in the $n$-node network. Here, we illustrate the methodology for constructing chiral-flow Hamiltonians from the boson-aSGF model by considering the examples of five- and six-node networks, corresponding to odd and even values of $n$, respectively, where the initial condition is still denoted by an excitation localized on the first node. The Hamiltonian realizing chiral flow in the $n$-node ($n>4$) network is labeled by $H_\mathrm{ch}^{n}$, as it further involves NNN hoppings in comparison to the simple aSGF Hamiltonian $H_\mathrm{aSGF}^n$. We first consider the five-node case, a paradigm for odd $n$. Since the eigenstates are independent of the coupling strength $\beta_c$, the excitation amplitude for $j$th node is given by
\begin{equation}
C_e^j(t)=\sum_{m=1}^{2} \frac{2}{5}\cos\left[E_m t-\frac{2(j-1)m\pi}{5}\right]+\frac{1}{5}\cos(E_{\beta_c}t).
\end{equation}
Therefore, conducting a perfect chiral flow requires the complete transmission condition: $\max[|C_e^j(t)|^2]=1$ for all sites, leading to the conditions $E_2=2E_1$ and 
\begin{equation}
\cos\frac{2\pi}{5}+\cos\frac{4\pi}{5}+\frac{1}{2}\cos\frac{2\pi E_{\beta_c}}{5E_1}=0,
\end{equation}
which also reduces to $E_{\beta_c}=5E_1$. According to the eigenstates and these constraints on eigenenergies, we yield the following chiral-flow Hamiltonian from this five-node boson-aSGF model, with NNN hopping:
\begin{equation}
H_\mathrm{ch}^5/J_0=\sum_{j=1}^5 -i(a_j^\dagger a_{j+1} - \alpha a_j^\dagger a_{j+2}) - \beta_c a_j^\dagger c + \text{H.c.},
\end{equation}
where $\alpha=\sqrt{(3-\sqrt{5})/2},~\beta_c=5\sqrt{2/(5+\sqrt{5})}$. In Fig.~\ref{Fig3}(a), we plot the dynamical excitation flow with Hamiltonian $H_\mathrm{ch}^5$. Here, for a clockwise chiral flow, the NN hopping phase is $-\pi/2$ instead of $\pi/2$. Both the NN and NNN hoppings exhibit a phase of $(p+1/2)\pi$ with $p\in\mathbb{Z}$ to preserve the chiral symmetry.

Similarly, from the eigenstates of the six-node network, we make the ansatz of the probability amplitude without giving energy levels,
\begin{equation}
\label{6nodeaSGFCej}
\begin{split}
C_e^j(t)=&(-1)^{(j-1)}\left\{\sum_{m'=1}^2\frac{1}{3}\cos\left[E_{m'} t-\frac{(j-1)m'\pi}{3}\right]\right.\\
&\qquad + \left.\frac{1}{6}\cos\left[E_{\beta_c}t-(j-1)\pi\right]+\frac{1}{6}\right\}\\
=& \sum_{m=1}^2\frac{1}{3}\cos\left[E_{3-m} t+ \frac{(j-1)m\pi}{3}\right]\\
&\qquad +\frac{1}{6}\cos E_{\beta_c}t+\frac{(-1)^{j-1}}{6},
\end{split}
\end{equation}
where the factor $(-1)^{j-1}$ is interpreted as $\cos(j-1)\pi$. Hence, the eigenstates $|\varphi_{\pm m}\rangle$ correspond to energies $\mp E_{3-m}$. From Eq.~\eqref{6nodeaSGFCej}, we can derive the energy-level relationships as follows: $E_1=2E_2$ and $E_{\beta_c}=3E_2$. With the eigenstates and energy levels, we obtain the six-node chiral-flow Hamiltonian
\begin{equation}
H_\mathrm{ch}^6/J_0 = \sum_{j=1}^6 i(a_j^\dagger a_{j+1} + \frac{1}{3}a^\dagger a_{j+2})-\sqrt{2} a_j^\dagger c + \text{H.c.}.
\end{equation}
In Fig.~\ref{Fig3}(b), we show the perfect chiral flow in the six-node network assisted by the auxiliary node. The six-node aSGF network with perfect chiral flow exhibits a total flux of zero, while breaking TR symmetry in its subtriangle structure. This is analogous to a plaquette in the Haldane model~\cite{haldane1988model}, a model that demonstrates chiral edge currents without magnetic flux.

By investigating networks with odd and even numbers of nodes, we have found a universal scheme for implementing the chiral flow within both odd and even categories of $n$. When we go to the $n$-node boson-aSGF network, there are more interactions involved, with the Hamiltonian having the form
\begin{equation}
\label{nnodeaSGF}
H_\mathrm{ch}^n/J_0 =\sum_{j=1}^n\sum_{k=1}^{\lceil n/2\rceil-1}i\alpha_k a_j^\dagger a_{j+k} + \beta_c a_j^\dagger c + \mathrm{H.c.},
\end{equation}
where $\lceil \cdot\rceil$ is the ceiling function, and $\alpha_k$ denotes the strengths of different kinds of couplings between nodes. Since $H_{\mathrm{ch}}^n$ shares the same chiral modes with $H_{\mathrm{aSGF}}^n$, it is assumed to have the same form as the Hamiltonian in Eq.~\eqref{equ:intermediate_diag}, with the only difference lying in the spectrum $E_m$; however, as the number of nodes $n$ increases, the complexity of the aSGF network for conducting a perfect chiral flow also increases. Specifically, the number of required couplings exhibits a scaling of $n^2/2$, which results in it being difficult to construct a chiral flow in a practical system. Thus, we will turn to finding an alternative way to realize chiral transmission between remote nodes in a large network in Sec.~\ref{Sec:MNN}.

\subsection{Spin-aSGF model and multiexcitation}

% ---------------------------------------------------------------------------
\begin{figure}[tbp]
\centering
\includegraphics[width=1.0\linewidth]{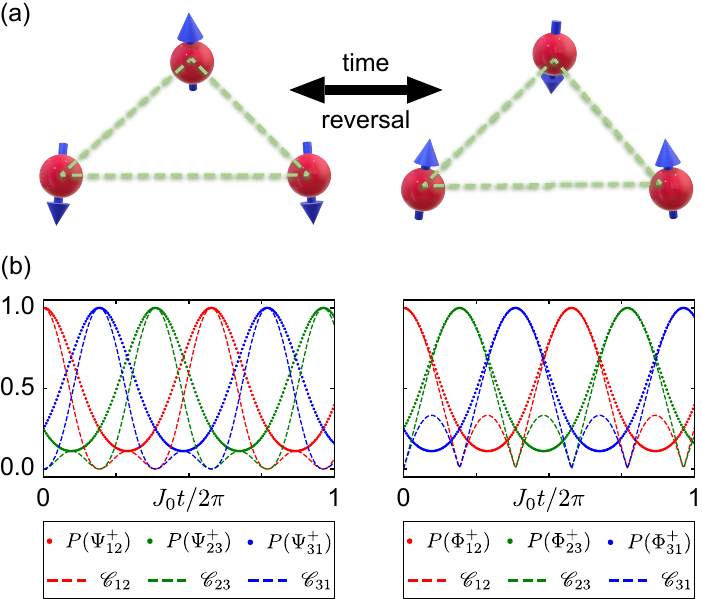}
\caption{(a) Schematic of the three-node spin-SGF model. TR symmetry predicts opposite chiral flows for the spin-up and spin-down states. (b) The time evolution of Bell states when the hopping phase is $\pi/2$. The Bell states $|\Psi_{jk}^+\rangle$ and $|\Phi_{jk}^+\rangle$ corresponding to different excitation subspaces have opposite chiral dynamics, where the concurrence of two spins (or atoms) $\mathscr{C}_{jk}$ transfers with the evolution of population $P(|\Psi_{jk}^+\rangle)$ or $P(|\Phi_{jk}^+\rangle)$.}
\label{Fig8}
\end{figure}
% ---------------------------------------------------------------------------

In the two-level atomic and spin systems, we can also develop the spin-SGF model by synthesizing the complex hoppings between atoms~\cite{Phys.Rev.Lett.2021DeBernardis,Phys.Rev.Research2022Wu} or the spin interaction~\cite{Nat.Phys.2019Wang}; however, the spin-SGF model exhibits very different symmetries compared to the boson-SGF model, resulting in distinctive chiral dynamics. By replacing the boson annihilation and creation operators $a_j$ and $a_j^\dagger$ with ladder operators $\sigma_j^-$ and $\sigma_j^+$, we obtain the spin-SGF Hamiltonian
\begin{equation}
H_\mathrm{SGF}^n = J_0 \sum_{\langle j k\rangle}^n (e^{i\theta_{jk}} \sigma_j^+ \sigma_k^- + \mathrm{H.c.}).
\end{equation}
This tight-binding Hamiltonian can describe both the excitation hopping between two-level atoms and the spin-exchange interaction. In contrast to the description in spinless lattice models~\cite{Phys.Rev.A2010Koch}, the TR operator for the spin system is defined as $\hat{T}=i\sigma_y \mathcal{K}$, where $\mathcal{K}$ denotes the complex conjugation. The operator $\hat{T}$ flips the spin vector as
\begin{equation}
\hat{T} \boldsymbol{\sigma} \hat{T}^{-1}=-\boldsymbol{\sigma}.
\end{equation}
Under the TR operation, the hopping term transforms to its Hermitian conjugate as $\hat{T} e^{i\theta_{jk}} \sigma_j^+ \sigma_k^- \hat{T}^{-1} = e^{-i\theta_{jk}} \sigma_k^+ \sigma_j^-$. Thus, the spin-SGF Hamiltonian is always TR invariant regardless of the specific values of the hopping phases $\theta_{jk}$. 

In the three-node spin-SGF network, when we set the hopping phases as $\theta_{jk}=\pi/2$, the system also supports a perfect chiral flow for the spin-up state $\left|\uparrow\downarrow\downarrow\right\rangle$ like the boson-SGF case. Nonetheless, the system's TR symmetry guarantees an opposite chiral evolution of the spin-down state $\left|\downarrow\uparrow\uparrow\right\rangle$, since it represents the TR counterpart of the spin-up state, as shown in Fig.~\ref{Fig8}(a). Analogous to the quantum spin Hall effect, the opposite chiral spin currents show the presence of TR symmetry, whereas in boson-SGF networks, the chiral flows are permitted in a single direction. 

Furthermore, based on the unique dynamics in the spin-SGF network, we can establish opposite chiral transmission channels for different entanglement states. In Fig.~\ref{Fig8}(b), we illustrate the opposite chiral dynamics of two types of Bell states $|\Psi^+_{jk}\rangle=(\left|\uparrow_j\downarrow_k\right\rangle + \left|\downarrow_j\uparrow_k\right\rangle)/\sqrt{2}$ and $|\Phi^+_{jk}\rangle=(\left|\downarrow_j\downarrow_k\right\rangle + \left|\uparrow_j\uparrow_k\right\rangle)/\sqrt{2}$, which correspond to the spin-up and spin-down chiralities. Here, we set the initial states as $|\Psi_{12}^+\rangle\otimes\left|\downarrow_3\right\rangle$ and $|\Phi_{12}^+\rangle\otimes\left|\downarrow_3\right\rangle$ in the left- and right-hand panels of Fig.~\ref{Fig8}(b), respectively. Both the populations of the Bell states and the concurrence $\mathscr{C}_{jk}$ can characterize the chirality of entanglement in this spin network. Because the dynamics of the two Bell states are constrained in the up and down subspaces, respectively, the preserved TR symmetry leads to opposite flow directions for them and causes distinctive entanglement flows; however, in the bosonic network with broken TR symmetry, the chirality remains consistent for different excitation subspaces. Hence, the transmission of entanglement is not sensitive to its initial state. Notably, the three-node case of the spin-SGF Hamiltonian, with the $\pi/2$ hopping phase, is also equivalent to the chiral antisymmetric spin exchange interaction (ASI) Hamiltonian~\cite{J.Phys.Chem.Solids1958Dzyaloshinsky,Phys.Rev.Lett.1960Moriya}, which breaks parity symmetry but preserves TR symmetry~\cite{Nat.Phys.2019Wang}.

% ---------------------------------------------------------------------------
\begin{figure}[tbp]
\centering
\includegraphics[width=1.0\linewidth]{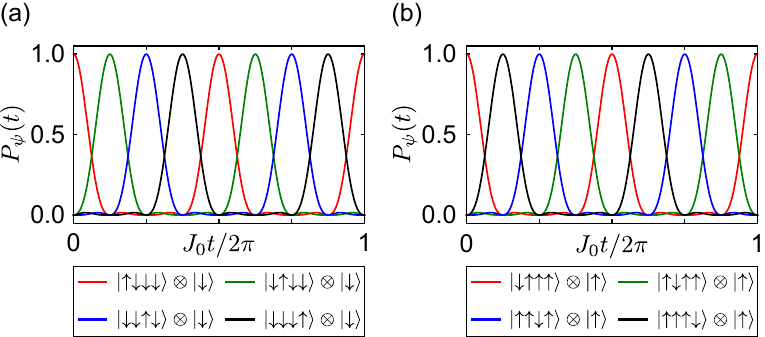}
\caption{Time evolution of the four-node spin-aSGF model. The spin-up and spin-down states of network nodes show opposite chirality with the assistance of auxiliary spin.}
\label{Fig9}
\end{figure}
% ---------------------------------------------------------------------------
\begin{table*}[tbp]
\caption{Review of different models supporting the three-node chiral flow.}
\label{TAB1}
\centering
\begin{ruledtabular}
\begin{tabular}{ccccc}
model & boson-SGF &  \multicolumn{2}{c}{ASI (spin-SGF)} & SCI \\
\midrule
Hamiltonian & $\sum_{j=1}^3 J_0(e^{i\pi/2}a_j^{\dagger}a_{j+1}+\mathrm{H.c.})$  & \multicolumn{2}{c}{$\hbar \kappa C_z S_z,~S_z=\sum_{j=1}^3\sigma_j^z/2$} & $\hbar \kappa C_z=\hbar\kappa\boldsymbol\sigma_1\cdot(\boldsymbol\sigma_2\times\boldsymbol\sigma_3)$\\
PT symmetry & $\hat{T}$-broken  & \multicolumn{2}{c}{$\hat{P}$-broken, $\hat{T}$-preserved} & $\hat{T}$-broken, $\hat{P}$-broken \\
Chiral symmetry & \checkmark & \multicolumn{2}{c} \checkmark & \checkmark
\\
\cmidrule { 3 - 4 } 
Subspace & Single-excitation & Single-excitation & Double-excitation & Single/double-excitation \\ 
$\begin{array}{c}
j\text {-excitation} \\
\text{Hamiltonian }\mathcal{H}^{(j)}
\end{array}$ & $\mathcal{H}^{(1)}\sim\left(\begin{array}{ccc}
0 & i & -i  \\
-i & 0 & i  \\
i & -i & 0
\end{array}\right)$ & $\mathcal{H}^{(1)}\sim\left(\begin{array}{ccc}
0 & i & -i  \\
-i & 0 & i  \\
i & -i & 0
\end{array}\right)$ & $\mathcal{H}^{(2)}\sim\left(\begin{array}{ccc}
0 & -i & i  \\
i & 0 & -i  \\
-i & i & 0
\end{array}\right)$ & $\mathcal{H}^{(1/2)}\sim\left(\begin{array}{ccc}
0 & -i & i  \\
i & 0 & -i  \\
-i & i & 0
\end{array}\right)$\\
Flow direction & $1\rightarrow3\rightarrow2\rightarrow1$ & $1\rightarrow3\rightarrow2\rightarrow1$ & $1\rightarrow2\rightarrow3\rightarrow1$ & $1\rightarrow2\rightarrow3\rightarrow1$ \\
Symmetric spectrum & \checkmark & \checkmark & \checkmark & \checkmark \\
\end{tabular}
\end{ruledtabular}
\end{table*}
% ---------------------------------------------------------------------------

Similarly, the $n$-node chiral flow can also be created in a spin or atomic system by employing the aSGF model by replacing the operators $a^{(\dagger)} \rightarrow \sigma^{-(+)}$ in Eq.~\eqref{nnodeaSGF}. As shown in Fig.~\ref{Fig9}(a), we observe chiral dynamics of the spin-up state in the four-node spin-aSGF network. If one then flips all the spins, an opposite chiral flow will arise between the network nodes, predicted by TR symmetry, as shown in Fig.~\ref{Fig9}(b); however, in contrast to the bosonic system, the spin-aSGF model does not support multiexcitation chiral flows and only permits the chiral dynamics for a single up or down state. In boson-aSGF networks with multiexcitation, the unidirectional chiral dynamics exists in each excitation subspace. Since the Hamiltonian does not contain photon-photon interactions, these excitations do not interact with each other and will display the same chiral motion irrespective of the excitation number. Therefore, it can be used to transfer or prepare more complicated entangled states~\cite{Nat.Phys.2019Wang}, such as the Greenberger-Horne-Zeilinger state and the NOON state~\cite{Phys.Rev.A2022Qi}.

\subsection{Discussion of the criteria}

From the study of constructing the Hamiltonian~\eqref{nnodeaSGF} for achieving the $n$-node chiral flow, we conclude that there are two essential ingredients for inducing perfect chiral flows: (i) a symmetric and equally spaced energy spectrum in a specific excitation subspace and (ii) a complete set of the paired chiral modes. The symmetric spectrum is usually obtained from the chiral symmetry of Hamiltonians in previous models, and it guarantees that each set of paired chiral modes with opposite energies contributes a propagating wave of the excitation in the network. The equally spaced energy levels then facilitate the interference of all pairs of chiral modes on each site, leading to a perfect site-by-site chiral flow. In particular, chiral symmetry of the whole Hamiltonian is not a prerequisite because we can build a symmetric spectrum and perform the perfect chiral flow exclusively within a specific subspace or several subspaces. For example, when we add the on-site potential $U$ (photon-photon interactions) to the three-node bosonic chiral-flow Hamiltonian such that $H'=\sum_{j=1}^3 U \hat{n}_j^2+ (iJ_0 a_j^\dagger a_{j+1} + \mathrm{H.c.})$, the chiral symmetry is broken; however, there is still a symmetric spectrum in the single-excitation subspace, allowing for the existence of the chiral flow. As the potential $U\gg J_0$, the energy spectrum in the double-excitation subspace also becomes nearly symmetric. In this sense, it forms a hard-core bosonic system and mimics the three-node spin-SGF network, resulting in opposite chiral flows in these two subspaces~\cite{Nat.Phys.2017Roushan}. 

Additional symmetries can also impose constraints on the chiral-flow Hamiltonian and produce previously unseen phenomena, e.g., the TR symmetry of the three-node spin-SGF model promotes different chirality, like the quantum spin Hall effect. In Table~\ref{TAB1}, we summarize different three-node models that support perfect chiral flows, and all of them manifest the chiral symmetry mentioned above. These models may break other symmetries within their systems, such as parity symmetry~\cite{Nat.Phys.2019Wang} (described by $\hat{P}$ operator) and TR symmetry, while the spin-chirality interaction (SCI) Hamiltonian~\cite{wen1989chiral,grohol2005spin,Appl.Phys.Lett.2020Liu} breaks both of them. In particular, all the Hamiltonians of the listed models have  $U(1)$ symmetry, which preserves the excitation number. Therefore, we can block-diagonalize the Hamiltonians as $H=\mathcal{H}^{(0)}\otimes\mathcal{H}^{(1)}\otimes\cdots\otimes\mathcal{H}^{(n)}$, where $\mathcal{H}^{(j)}$ is the block Hamiltonian in the $j$-excitation subspace (for a spin system the excitation is replaced by the spin-up state). For each model listed in Table 1, the symmetric energy spectrum of the block Hamiltonian $\mathcal{H}^{(j)}$ makes it possible for chiral flows to be present in the $j$th excitation subspace. In our aSGF model, the chiral symmetry is still preserved after introducing an auxiliary node if the hopping (including long-range hoppings in the $n$-node case) phase between network nodes is $\theta_{j,j+k}=(p+1/2)\pi$ with $p\in \mathbb{Z}$. The auxiliary node also generates the last required pair of chiral modes, which facilitates the completeness of chiral modes and thus restores perfect chiral flows within $n$-node aSGF networks.

% ---------------------------------------------------------------------------
% ---------------------------------------------------------------------------

\section{Extension of four-node networks}
\label{Sec:MNN}

\begin{figure}[tbp]
\centering
\includegraphics[width=1.0\linewidth]{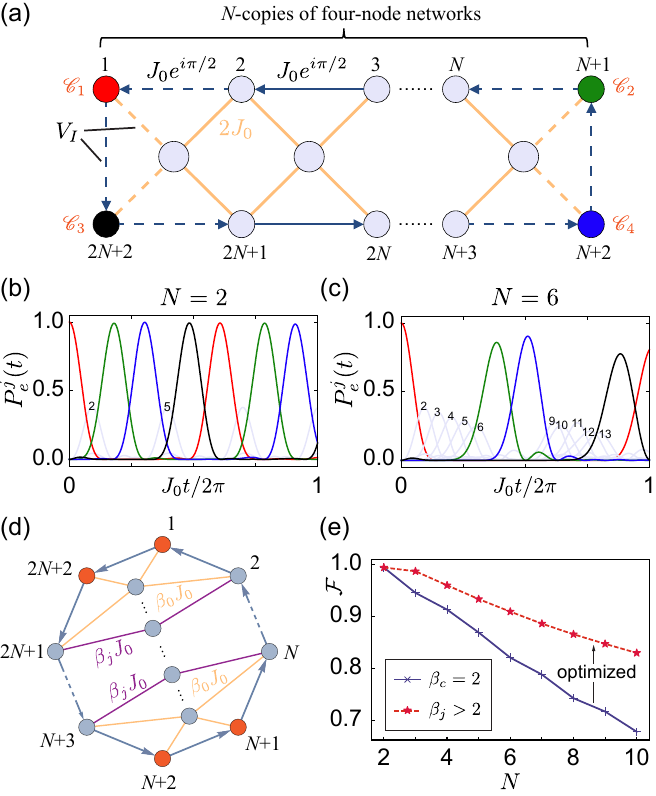}
\caption{(a) Illustration of the ladder model with auxiliary nodes, where the corner nodes are labeled as $\mathscr{C}_j (j=1,2,3,4)$. The solid and dashed arrows are hoppings partitioned into $H_0$ and $V_I$, respectively. (b),(c) Chiral flows with different copies of the four-node network($N=2$ and $N=6$ respectively). (d) Effective configuration of the ladder network. In this pattern, the corner nodes are equivalent. (e) Fidelity of chiral flow in the ladder network as a function of $N$. Here, the blue solid line represents the fidelity from identical coupling $\beta_c=2$, while the red dashed line denotes the results from optimal parameters with $\beta_j>2$ ($j=1,\cdots,m$).}
\label{Fig5}
\end{figure}

In this section, we extend the four-node chiral flow by connecting four-node networks to form a ladder. As depicted in Fig.~\ref{Fig5}(a), we arrange $N$ copies of the four-node boson-aSGF network alongside one another. Every copy shares two nodes with its left and right neighboring copies, and we have eliminated the hoppings between these shared nodes, e.g., node 2 and node $2N+1$. The hopping strengths and phases are set to be identical in all copies, and the Hamiltonian reads
\begin{equation}
    \label{equ:Ncopies}
    \begin{aligned}
    H_\mathrm{ex}^{N}/J_{0}&=\sum_{j=1}^{2N+2}\left(e^{i\theta_{0}}a_{j}^\dagger a_{j+1}+{\rm H.c.}\right)\\
        &\qquad +\beta_{c}\sum_{i=1}^{N}\sum_{j\in{\rm copy}i}\left(c_{i}^\dagger a_j+{\rm H.c.}\right),
    \end{aligned}
\end{equation}
where the relevant parameters are based on the four-node case. The index $j$ in the second term runs over the nodes belonging to copy $i$ of the four-node network. Each node, except the corners, belongs to two copies. This ladder network bears some resemblance to quasi-1D topological insulators such as the rhombus chain~\cite{vidal2000interaction} and the Creutz ladder~\cite{creutz1999end}. All of the hopping phases between nearest-neighboring nodes are set to be $\theta_0=\pi/2$ to ensure the chiral symmetry of the system, and the coupling strengths between auxiliary nodes and network nodes are the same as the coupling in the four-node case, i.e., $\beta_c=2$; however, this configuration cannot be interpreted as a synthetic uniform magnetic field model because the hopping phases in this structure cannot be the same within a uniform magnetic field. In practice, it can still be synthesized in superconducting circuits by accurately setting the Floquet modulation parameters~\cite{SciPostPhys.Lect.Notes2022Clerk} (see Sec.~\ref{Sec:sc_circuit} for details). 

% ---------------------------------------------------------------------------
% ---------------------------------------------------------------------------

We study the chiral dynamics of this ladder network by initiating an excitation at the first node. Figures~\ref{Fig5}(b) and \ref{Fig5}(c) display the evolution of the excited-state populations of all nodes for the $N=2$ and $N=6$ cases respectively, demonstrating that the primary characteristic of the chiral flow is retained. The majority of the excitation is chirally transported between the corner nodes $\mathscr{C}_j$ [as labeled in Fig.~\ref{Fig5}(a)], namely, $1\rightarrow N+1\rightarrow N+2\rightarrow 2N+2$. We emphasize that the imperfection of this chiral flow originates from the unevenly spaced oscillation frequencies, as detailed in Appendix~\ref{ExtendedNetwork}, where we also provide the analytic expression for $P_e^1(t)$ by partitioning the Hamiltonian as $H_\mathrm{ex}^{N}=H_0+V_I$ and implementing the nonperturbative calculation of system's propagator. As the number of copies $N$ increases, the timescale for transferring an excitation from node $1$ to node $N+1$ also increases in a linear form (see Appendix~\ref{Appendix:OLN}). To describe the transmission rate of this chiral flow, we define the fidelity as $\mathcal{F}=|\langle\psi(t=0)|\psi_f\rangle|^2$, where $\psi_f$ is the final state after a cycle period. As shown in Fig.~\ref{Fig5}(e), the fidelity of chiral flow $\mathcal{F}$ decays linearly when the network becomes larger, with less remaining excitation at the first node.

The extended four-node model can also be viewed as a polygonal network with auxiliary nodes, as shown in Fig.~\ref{Fig5}(d). Within this pattern, we can still maintain equal treatment of the corner nodes by replacing $\beta_c$ with a sequence of coupling parameters $\beta_j (j=0,1,\cdots m$, where $m=\lceil N/2 \rceil -1$). Thus, we can then optimize the network by modulating the parameters. Intuitively, the coupling $\beta_j J_0$ should be enhanced to connect further nodes when it becomes distant from the corner nodes, i.e., $2=\beta_0<\beta_1<\cdots<\beta_m$.  This has been verified by numerical calculation, where we systematically iterate through these parameters to identify the optimal values (see Appendix~\ref{Appendix:OLN}). In Fig.~\ref{Fig5}(e), we plot the optimized fidelity as a function of $N$ with optimal parameters, which shows a linear decline but is improved measurably compared to the identical coupling case. Even if the network expands to ten times the four-node network's scale, the fidelity remains at $\mathcal{F}>0.8$. Therefore, we are convinced that the chirality persists and dominates within a larger network, rendering this ladder network highly practical for implementing chiral transmission between remote nodes.

% ---------------------------------------------------------------------------
% ---------------------------------------------------------------------------

\section{Imperfections in networks}\label{Sec:IMP}

In practical systems, the inevitable imperfections from various aspects of quantum networks can affect the quality of chiral flows in the $n$-node aSGF network and ladder networks. To estimate such influence, in this section, we investigate the sensitivity of chiral flow to the disorder on the node (bosonic mode or atomic transition) frequency $\omega_0$, coupling strength $J_0$, and hopping phase $\theta_{ij}$. Here, we take the four-node aSGF model as an example for illustration. 

\begin{figure}[tbp]
\centering
\includegraphics[width=1.0\linewidth]{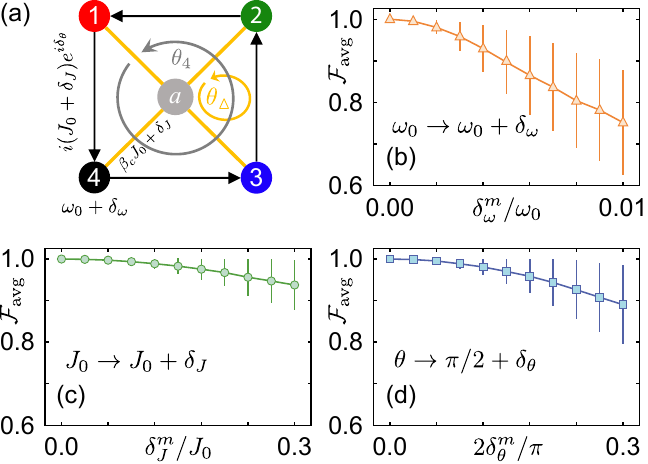}
\caption{(a) Illustration of the four-node aSGF model with inevitable imperfections. The imperfections of this network are described by the disorder of the node frequencies and hoppings. (b)-(d) Effects of different types of disorder on the average fidelity. The numerical data represent the average values derived from 2000 instances of disorder. The error bars in the graph signify one standard deviation of the data points.}
\label{Fig4}
\end{figure}

As shown in Fig.~\ref{Fig4}(a), these three types of disorder are denoted as $\delta_\square$, $\square = \omega, J$, and $\theta$, where the first acts on the node frequency while the latter two affect the hoppings between nodes. In the single-excitation subspace, the disorder on frequency and hopping are represented as the diagonal and off-diagonal disorder of the single-particle Hamiltonian $\mathcal{H}_\mathrm{aSGF}^{4}$, respectively. We first consider the disorder on the frequency by replacing $\omega_0$ with $\omega_0 + \delta_\omega$ for each node, where $\delta_\omega$ is sampled from a uniform distribution within the range [$-\delta_{\omega}^m,\delta_{\omega}^m$]. Similarly, the coupling strengths and hopping phases are sampled among $[J_0-\delta^m_J,J_0+\delta^m_J]$ and $[\pi/2-\delta^m_\theta,\pi/2+\delta^m_\theta]$, respectively. In our simulations, the parameters are taken from experimental data based on a superconducting circuit platform~\cite{Nat.Phys.2019Wang}, where transmon qubits serve as artificial atoms and complex hopping terms are implemented via a superconducting bus resonator (see Sec.~\ref{Sec:sc_circuit} for details). In such a platform, imperfections primarily arise from fabrication-induced disorder, unintended crosstalk between closely spaced circuit elements, and material defects, all of which collectively lead to deviations from the designed parameters. Specifically, we use $\omega_0 = 5.6~\mathrm{GHz}$, which represents the frequency of each artificial atom, and $J_0 = 2\pi \times 4.29~\mathrm{MHz}$, which denotes the hopping strength.
Since the maximum excitation population of the $j$th node is no longer 1, i.e., $\max\left[P_e^j(t)\right]<1$ within a typical cycle period, we use the average fidelity of the corner excitation to characterize the quality of chiral flow: 
\begin{equation}
\mathcal{F}_\mathrm{avg} = \sum_{j=1}^4 \max\left[C_e^j (t)\right]/4.
\end{equation}
As illustrated in Fig.~\ref{Fig4}(b), the frequency disorder significantly influences the chiral flow.  Compared with the hopping, even a slight deviation of the node frequency can greatly change the energy spectrum, leading to a considerable decrease in the fidelity $\mathcal{F}_\mathrm{avg}$. Moreover, the chiral symmetry is broken when such diagonal (i.e., frequency) disorder is introduced, which results in asymmetry of the energy spectrum and thus violates the criteria for conducting a perfect chiral flow. In contrast, off-diagonal disorder preserves the chiral symmetry of the system and has only minor impacts on the energy spectrum. As plotted in Figs.~\ref{Fig4}(c) and \ref{Fig4}(d), the chiral flow shows robustness ($\mathcal{F}_\mathrm{avg}>0.9$ with about $30\%$ deviation of the parameters) when the disorder stems from the hopping strength and phase.
% ---------------------------------------------------------------------------
% ---------------------------------------------------------------------------
% ---------------------------------------------------------------------------

\section{Feasibility in experiments}
\label{Sec:Exper}

In this section, we elucidate the practical implementation of our proposal, where we take the four-node aSGF model as an example in the discussion that follows. Here, we first briefly introduce the schemes for superconducting circuits~\cite{Nat.Phys.2017Roushan,Nat.Phys.2019Wang} and magnetic photonic lattice platforms~\cite{NaturePhoton2012Fang,NaturePhoton2013Hafezi,PhysRevX.4.031031,Phys.Rev.Lett.2021DeBernardis} to achieve controllable hoppings with phases, thus promising the realization of multinode chiral flows in bosonic and atomic systems, respectively. We then elaborate on the synthesis of gauge fields in ultracold atoms, which offers another promising platform for multinode chiral flows in both spin and bosonic systems.

\subsection{Superconducting circuits}\label{Sec:sc_circuit}

The hopping phases between the nodes of superconducting setups, such as resonators and qubits~\cite{you2011atomic,devoret2013superconducting,Wendin_2017,Blais2020,blais2021circuit}, can be synthesized in two ways: (1) by introducing tunable couplers to periodically modulate coupling strengths~\cite{Nat.Phys.2017Roushan,kapit2013quantum,Phys.Rev.Lett.2014Chen}, or (2) by periodically adjusting the transition frequencies of nodes, which are coupled to a mutual bus resonator~\cite{Nat.Phys.2019Wang,Phys.Rev.X2014Goldman}.

Following the first approach, we can design a superconducting circuit consisting of five transmon qubits (or resonators) $Q_1$, $Q_2$, $Q_3$, $Q_4$ and $Q_{a}$, coupled with each other via tunable couplers, where the couplings can be dynamically modulated as $g_{jk}(t)=2g_{jk}\cos\left(\nu_{jk}t+\phi_{jk}\right)$~\cite{Phys.Rev.Lett.2014Chen}. Here, the modulation frequency $\nu_{jk}$ is chosen to be the frequency detuning between elements $j$ and $k$, i.e., $\nu_{jk}=\omega_j-\omega_k$, and we assume that $\nu_{jk}\gg g$. After transforming to the interaction picture under the rotating-wave approximation (RWA), we obtain the effective Hamiltonian
\begin{equation}\label{equ:4+1model}
    \begin{aligned}
        H_{{\rm eff}}=&\sum_{\substack{j,k=1\\j<k}}^{4}\left(g_{jk}e^{-i\phi_{jk}}a_{j}^\dagger a_{k}+{\rm H.c.}\right)\\
    &+\sum_{j=1}^{4}\left(g_{j}^c e^{-i\phi_{j}^c} a_{j}^\dagger c+{\rm H.c.}\right)+ H_\mathrm{nl},
    \end{aligned}
\end{equation}
where $H_\mathrm{nl}\approx -U/2\left(\sum_j a_j^\dagger a_j^\dagger a_j a_j + c^\dagger c^\dagger c c\right)$ is the Kerr nonlinearity and arises from the cosine potential of a transmon qubit, with the on-site potential $U\gg g$. The nonlinearity of superconducting qubits naturally promotes the effective Hamiltonian as a spin-aSGF model since it prohibits the double-photon occupancy if the transmons are initially singly occupied with double excitation denoted
by $a_j^\dagger a_k^\dagger |G\rangle$. For harmonic resonators with $U=0$, $H_\mathrm{eff}$ mimics a boson-SGF Hamiltonian. To satisfy the requirements of conducting a four-node chiral flow, we set the parameters as follows: $g_{j}^c/2=g_{j,j+1\mathrm{mod}4}=g$ for $j=1$, 2, 3, and 4; $g_{13}=g_{24}=0$; $\phi_{j}^c=0$; and $\phi_{j,j+1\mathrm{mod}4}=-\pi/2$. Notably, the frequency detunings of the elements have to satisfy the RWA conditions: $|\omega_j-\omega_k|>g$ for $j\neq k$. The Hamiltonian then reduces to the form in Eq.~\eqref{equ:SGFCmodel} with $N=4$, $J_0=g$, and $\beta_c=2$, allowing for the conduction of the perfect chiral flow.

In addition, the first sum of Hamiltonian~\eqref{equ:4+1model} can be synthesized via a bus resonator~\cite{Nat.Phys.2019Wang}. In this circuit quantum electrodynamics (QED) model, four elements are interconnected by a bus resonator with couplings $g_{j}$. To construct synthetic gauge fields, we periodically adjust the frequencies of qubits around the frequency of the bus resonator $\omega_r$ as
\begin{equation}
\omega_{j}(t)=\omega_{r}+\Delta \cos \left(\nu t-\phi_{j}\right),
\end{equation}
where $\Delta$ and $\nu$ represent the modulation amplitude and frequency, with $\nu \approx \Delta/2.40$, and the phases are denoted as $\phi_j=j\pi/2$. Under the condition $\nu\gg g$, we obtain the effective first-order Hamiltonian with complex hopping amplitudes $g_{jk}={g_{j}g_{k}}\beta_{jk}e^{i\pi/2}/{\nu}$, by neglecting higher-order terms. Here, $\beta_{jk}$ is given by the expression $\beta_{jk}=\sum_{n=1}^{+\infty}2J_{n}^{2}\left[f\right]\sin\left[n\left(\phi_{k}-\phi_{j}\right)\right]/n$, where $J_n[x]$ represents the $n$th-order Bessel function of the first kind and the dimensionless variable $f=\Delta/\nu\approx2.40$~\cite{kyriienko2018floquet,wu2018efficient} such that the zeroth-order term $J_0[f]=0$. The $\pi/2$ hopping phase of $g_{jk}$ can thus accurately replicate the four-node chiral-flow Hamiltonian. By setting $\phi_j=j\pi/2$, we can establish NN hoppings with the same strength while preventing the occurrence of NNN hoppings between nodes $j$ and $j+2$.

\subsection{Emitters in magnetic photonic lattices}
The circuit QED provides a convenient way to synthesize hopping phases between artificial atoms or resonators. Here, we show that a hybrid system, i.e., a magnetic photonic lattice (MPL) coupled to atomic emitters~\cite{Phys.Rev.Lett.2021DeBernardis}, can also induce hopping phases and be used to construct our model. Although the chiral flow between emitters is imperfect, we still see this dominant feature in such a hybrid system.

\begin{figure}
    \centering
    \includegraphics[width=0.48\textwidth]{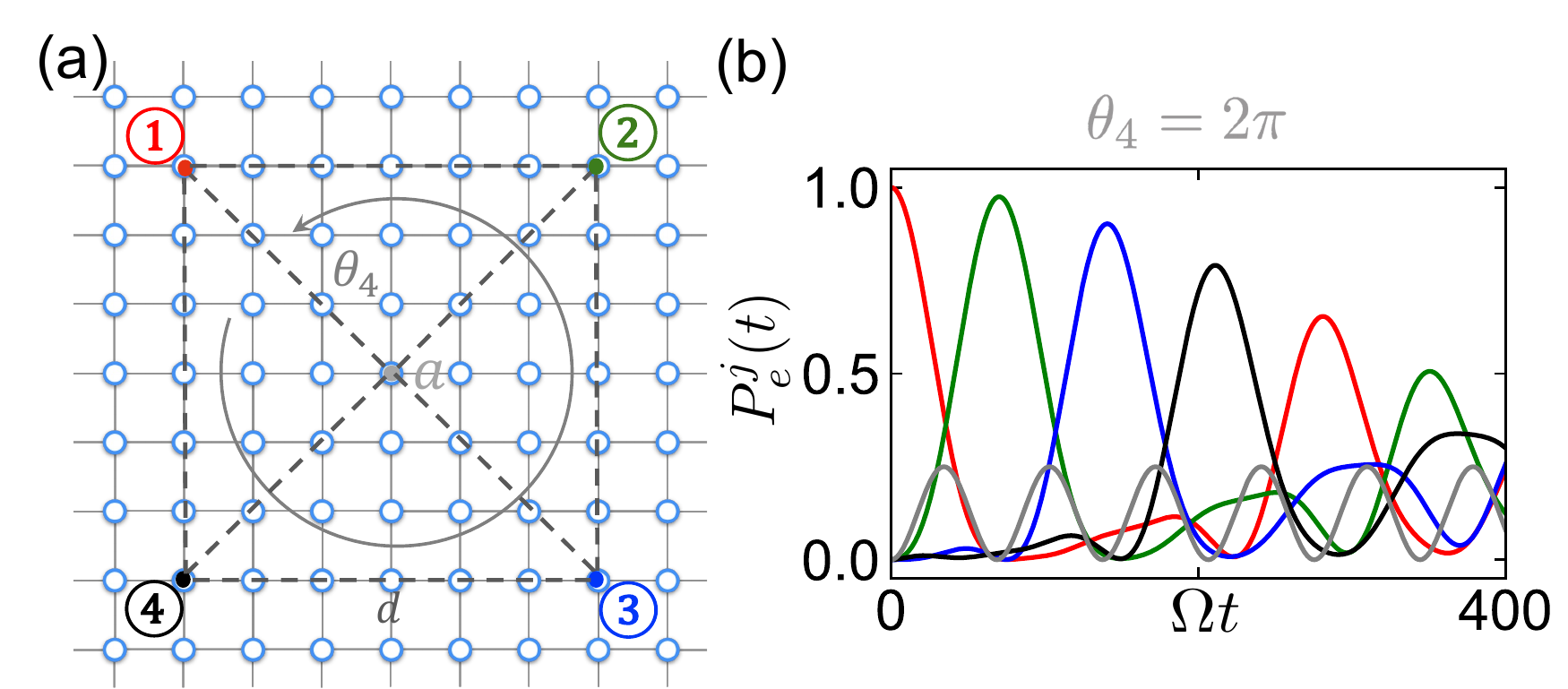}
    \caption{(a) Schematic of an aSGF model in a magnetic photonic lattice. The blue hollow circles are resonators that form a square lattice, while the filled circles with five different colors are emitters coupled to the resonators locally. The dashed lines between emitters indicate the effective long-range dipole-dipole interactions induced by the magnetic photonic lattice. (b) Evolution of the excited-state populations of emitters in the MPL, where $\ell=0$, $g/(\omega_e-\omega_{\ell=0})=0.1$, the distance between nearest emitter is $|\boldsymbol{r}_j-\boldsymbol{r}_{j+1}|/l_0=4$ and the unit of time is $\Omega^{-1}=(|\boldsymbol{r}_j-\boldsymbol{r}_{j+1}|/l_0)g^{-1}$.}
    \label{fig:MPL}
\end{figure}

As shown in Fig.~\ref{fig:MPL}(a), the MPL is a square array composed of identical photonic resonators with equal spacing $l_0$ and frequency $\omega_r$, where the neighboring lattices are coupled via a synthetic magnetic field $\boldsymbol{B}$ perpendicular to the array. The Hamiltonian of the MPL is a tight-binding lattice model with complex tunneling amplitudes. It is given by
\begin{equation}
H_{\mathrm{MPL}}=\omega_r\sum_{i}\Psi^\dagger_i\Psi_i-J\sum_{\langle i,j\rangle}(e^{i\theta_{ij}} \Psi^\dagger_i\Psi_{j}+\mathrm{H.c.}),
\end{equation}
where $\Psi^{(\dagger)}_i$ annihilates (creates) a photon of frequency $\omega_r$ inside the resonator at $\boldsymbol{r}_i$. In the continuum limit, this Hamiltonian can be approximately diagonalized into a series of Landau levels with energies $\omega_\ell \sim B J(\ell+1/2)$ ($\ell=0,1,2,\dots$), each of which possesses a large number of degenerate states~\cite{Phys.Rev.Lett.2021DeBernardis}.

We place five atomic emitters in resonators located at the four corners and the center of a square with side length $d$. These two-level emitters are described by the Pauli matrices $\sigma^j$ and are labeled as 1, 2, 3, 4, and $a$. They have the same emitter frequency $\omega_e$. Each emitter is locally coupled to its corresponding on-site resonator through a standard Jaynes-Cummings term, $g(\Psi_j \sigma_{+}^{j}+\Psi^{\dagger}_j \sigma_{-}^{j})$, where $g$ represents the uniform coupling strength for all five emitter-resonator pairs. The MPL acts as a structured photonic bath, mediating photon-exchange interactions among the emitters through collective photonic modes. If the coupling strength is much smaller than the system's Landau energy gap, i.e., $g\ll \omega_{\ell+1}-\omega_\ell$, the emitters are effectively coupled to the Landau level closest to $\omega_e$, which is assumed to be $\ell=0$. Under the condition of large detuning $|\omega_e-\omega_{\ell=0}|\gg g$, we can apply a Schrieffer-Wolff transformation to obtain the effective Hamiltonian
% ---------------------------------------------------------------------------
% --------------------------------------------------------------------------- 
\begin{equation}
H_{\rm eff}=\sum_{j,k}\tilde{J}_{jk}\sigma_+^j\sigma_-^k + \mathrm{H.c.},
\end{equation}
which contains only the long-range dipole-dipole interaction. The complex hopping amplitude is given by $\tilde{J}_{jk} = g^2 e^{i\theta _{jk}} G_0(\boldsymbol{r_j},\boldsymbol{r}_k)/(\omega_e - \omega_{\ell=0})$, where $G_0(\boldsymbol{r_j},\boldsymbol{r}_k)$ denotes the spatial component of the photonic Green’s function~\cite{Phys.Rev.Lett.2021DeBernardis}. By setting the gauge-invariant flux $\theta_4\equiv \theta_{12}+\theta_{23}+\theta_{34}+\theta_{41}=2\pi$ to break the TR symmetry of each subtriangle plaquette, this effective dipole-dipole interaction can to some extent simulate the four-node spin-aSGF model. Figure ~\ref{fig:MPL}(b) shows the chiral dynamics of the emitter excitation, where two factors contribute to the imperfection of this chiral flow. First, in this uniform magnetic photonic lattice, the coupling parameter between the auxiliary emitter and node emitter will be $\beta_{c}=1.48$ instead of $2$. Consequently, this deviation prevents the generation of a perfect four-node chiral flow. Second, the unavoidable NNN hopping between emitters from diagonal sites also reduces the fidelity of the chiral flow transmission. Nevertheless, by introducing external coupling channels~\cite{NaturePhoton2013Hafezi,Phys.Rev.X2015Peano,Optica2015Schmidt,Nature2017Lodahl} to eliminate these effects, it is still possible to realize a perfect chiral flow between emitters in this system.

% ---------------------------------------------------------------------------
% ---------------------------------------------------------------------------

\subsection{Ultracold atoms}
The synthetic gauge field can also be found in many experiments with neutral atomic gases. Here, we first describe the general proposal for synthesizing the hopping phase in an atom-light coupling system, where the induced complex couplings between two-level atoms can be used to simulate the spin-SGF model. Then, we introduce the schemes for generating synthetic gauge fields in cold-atom systems, which are quite different from the two-level-system case and allow us to create chiral flows of bosonic atoms.
The flexibility of these schemes for generating synthetic gauge fields between atoms also enables the promising implementation of aSGF models in atomic systems.

\subsubsection{External driving fields with two-level atoms}

Transitions between atomic levels are usually controlled by external driving fields. One can easily obtain the hopping of a single photon with a certain phase by applying an off-resonant monochromatic field on two separated atoms. For two atoms coupling with a laser field, the full Hamiltonian has the following form:
\begin{equation}\label{ultracold:intbyph}
H/\hbar=\omega_ca^\dagger a + \sum_{i=1,2}\omega_i \sigma_i^+\sigma_i^-
+\Omega(\sigma_i^+ ae^{i\phi_i}+\mathrm{H.c.}),
\end{equation}
where $\omega_i$ ($i=1,2$) and $\omega_c$ are the frequencies of the atoms and the laser field, respectively. The atom-light coupling strength $\Omega$ denotes the Rabi frequency, and $\phi_i$ is the coupling phase imparted by the applied field. Spatially separated atoms experience distinct coupling phases due to position-dependent variations in the driving laser field, resulting in a phase difference $\phi=\phi_1(\boldsymbol{r}_1)-\phi_2(\boldsymbol{r}_2)$. For simplicity, we assume that both atoms share the same transition frequency ($\omega_1=\omega_2=\omega_0$), detuned from the laser frequency by $\Delta=\omega_0-\omega_c$. In the large-detuning regime where $\Delta\gg \Omega$, the atoms are virtually excited by the driving field. In the interaction picture with $H_0=\hbar\omega_c(a^\dagger a+\sigma_1^+\sigma_1^-+\sigma_2^+\sigma_2^-)$ and $V=H-H_0$, the Hamiltonian becomes  
\begin{equation}
\begin{aligned}
\widetilde{H}/\hbar=&e^{iH_0 t}Ve^{-iH_0 t}\\
=&\sum_{i=1,2}\Delta\sigma_i^+\sigma_i^-+\Omega(\sigma_i^+ ae^{i\phi_i}+\mathrm{H.c.}).
\end{aligned}
\end{equation}
To obtain the effective Hamiltonian of the dynamics, we perform a Schieffer-Wolff transformation such that $\widetilde{H}_\mathrm{eff}=e^{S}\widetilde{H}e^{-S}$, with $S=\sum_i \frac{\Omega}{\Delta}(\sigma_i^-a^\dagger - \sigma_i^+ a)$. We can then eliminate the degree of freedom of the light, and the atoms are thus decoupled from the light field. The effective Hamiltonian is given by
\begin{equation}
\widetilde{H}_\mathrm{eff}/\hbar
=\sum_{i=1,2}(\Delta+\lambda)\sigma_i^+\sigma_i^- +\lambda e^{i\phi}(\sigma_1^+\sigma_2^-+\mathrm{H.c.}),
\end{equation}
where $\lambda=\Omega^2/\Delta$ denotes the coupling strength and Lamb shift of atoms due to a second-order (virtual photon-exchange) process from the off-resonant atom-light coupling. The effective Hamiltonian indicates a complex-valued atom-atom interaction from the dynamics. The phase attached to the field now appears in the photon-exchange interaction between atoms. In this sense, giving different dressing fields to any two neighboring atoms induces different phases of excitation hopping. For $N$ atoms with monochromatic
dressing, the system's Hamiltonian is given by $H=\sum_{k=1}^{N-1}g e^{i\theta_k}\sigma_k^+\sigma_{k+1}^-+\mathrm{H.c.}$ If atoms form a closed coupling loop, the total phase---i.e., the synthetic flux---becomes a nontrivial quantity, which is gauge invariant under any gauge transformations. Since the two-level atom is described by Pauli spin operators, this process can be exploited to create chiral flows of spin.

\subsubsection{Noninertial frame of reference and geometric gauge
potentials with bosonic atoms}

Constructing the boson-SGF model entails the hopping of bosonic atoms instead of excitation. For charge-neutral atoms, SGFs are obtained in a noninertial frame of reference, as in rotating systems~\cite{matthews1999vortices,madison2000vortex,abo2001observation,williams2010observation}, or by inducing geometric phases~\cite{dum1996gauge,visser1998geometric,juzeliunas2006light,Lin2009,spielman2009raman}. In these scenarios, the center-of-mass motion of the atoms acquires Peierls phases that are linked to the spatial gauge fields within the system. Consider an atom of mass $M$ trapped in a potential field $V(\boldsymbol{r})$, where $\boldsymbol{r}$ denotes the center-of-mass position vector. The atomic Hamiltonian is then expressed as
\begin{equation}
H_\mathrm{at}=\frac{\boldsymbol{p}^2}{2M}+V(\boldsymbol{r}).
\end{equation}
If the system rotates about the direction of a unit vector $\hat{\mathbf{n}}$ at a constant angular frequency $\boldsymbol{\mathcal{W}}=\mathcal{W}\hat{\mathbf{n}}$, the position vector in the spatially rotating frame becomes $\boldsymbol{r}\rightarrow\boldsymbol{r}'=\mathcal{D}_\mathbf{\hat{n}}(t)\boldsymbol{r}\mathcal{D}_\mathbf{\hat{n}}^\dagger(t)$, where $\mathcal{D}_\mathbf{\hat{n}}(t)=\mathrm{exp}{(-i\boldsymbol{\mathcal{W}}\cdot \boldsymbol{L}t/\hbar)}$ is the rotation operator and $\boldsymbol{L}=\boldsymbol{r}\times \boldsymbol{p}$ is the angular momentum operator. To obtain a time-independent Hamiltonian, we work in the 
rotating frame such that~\cite{Goldman_2014}
\begin{equation}
\begin{aligned}
H=&\mathcal{D}_\mathbf{\hat{n}}^\dagger H_\mathrm{at}(\boldsymbol{r'}(t)) \mathcal{D}_\mathbf{\hat{n}}-i\hbar \mathcal{D}_\mathbf{\hat{n}}^\dagger \frac{\partial}{\partial t}\mathcal{D}_{\hat{\mathbf{n}}}\\
=&\frac{(\boldsymbol{p}-\boldsymbol A)^2}{2M}+V(\boldsymbol r)+W_\mathrm{cen}(\boldsymbol r),
\end{aligned}
\end{equation}where $\boldsymbol A=M\boldsymbol{\mathcal{W}}\times\boldsymbol r$ is the effective vector potential for neutral atoms and $W_\mathrm{cen}=-\boldsymbol{A}^2/2M$ represents a centrifugal potential due to rotation. The latter term can be counteracted by a trapping potential created with Gaussian beams. Thus, the effective magnetic field is given by $\boldsymbol{B}=\nabla\times\boldsymbol{A}=2M\boldsymbol{\mathcal{W}}$, which is proportional to the rotational angular velocity. This shows that the SGF is introduced by exploiting the equivalence between the Lorentz force and the Coriolis force in a rotating frame of reference. With atoms tunneling between independently rotating traps, their motion induces Peierls phases $\theta_{jk}=\frac{1}{\hbar}\int_{\boldsymbol{r}_k}^{\boldsymbol{r}_j}\boldsymbol{A}\cdot d\boldsymbol{r}$, enabling the simulation of a boson-SGF network to realize chiral flows of ultracold bosonic atoms. Another paradigm for achieving SGFs in a noninertial frame of reference is the Floquet engineering of optical lattices~\cite{aidelsburger2011experimental,struck2012tunable,hauke2012non,Struck2013,baur2014dynamic}, which is often related to the preparation of topological states. In periodically driven optical lattices, the modulated potentials generate inertial forces acting on atoms and give rise to tunable gauge vector potentials. In the effective Hamiltonian, the atomic hopping phases between neighboring lattice sites are controlled by shaking parameters, allowing us to establish certain SGF configurations and realize chiral flows within optical lattices.

In addition to engineering synthetic gauge fields in noninertial reference frames, employing geometric gauge potentials offers an alternative approach. Crucially, the internal degrees of freedom of atomic states provide a tool for creating geometric phases. Considering the internal states of atoms, the full Hamiltonian is now written as
\begin{equation}
H_\mathrm{at}=\left[\frac{\boldsymbol{p}^2}{2M}+V(\boldsymbol{r})\right]\mathbb{I}+\mathbb{M}(\boldsymbol{r}),
\end{equation}
where $\mathbb{I}$ is the identity operator in the internal Hilbert space. The coupling term $\mathbb{M}(\boldsymbol{r})$ between atomic internal states and the external field characterizes the dynamics of the atom’s internal degrees of freedom, and its spatial dependence links internal states with the atomic position. The atomic state vector can be expanded in the dressed-state basis $|\chi_m\rangle$ as
\begin{equation}\label{statevec}
|\psi(\boldsymbol{r}, t)\rangle=\sum_{m=1}^N \phi_m(\boldsymbol{r}, t)\left|\chi_m(\boldsymbol{r})\right\rangle.
\end{equation}
In the basis of dressed states, the coupling term $\mathbb{M}(\boldsymbol{r})$ becomes a diagonal matrix $\hat{\varepsilon}$ with eigenenergies $\varepsilon_m$, and the Hamiltonian in this representation reads~\cite{dum1996gauge,Goldman_2014}
\begin{equation}
\hat{H}=\frac{(\boldsymbol{p}-\boldsymbol A)^2}{2 M}+V(\boldsymbol{r})+\hat{\varepsilon}.
\end{equation}
This implies that atomic external motion is affected by a geometric gauge field, whose matrix elements are given by $\langle \chi_n|\boldsymbol A|\chi_m\rangle=i\hbar\langle \chi_n|\nabla \chi_m\rangle$. The geometric potential arises from the interplay between atomic external (center-of-mass) dynamics and its internal degree of freedom. Since the dressed states vary with atomic position, the momentum operator that acts on these position-dependent states effectively couples the internal Hilbert space. Suppose that nonadiabatic transitions between dressed states are significantly suppressed due to the large energy gap; in this case, we can project the system's dynamics onto a certain atomic ground-state manifold. If an atom is initially prepared in the internal state $|\chi_1\rangle$, by substituting Eq.~\eqref{statevec} into the time-dependent Schrödinger equation, we obtain the equation of external motion~\cite{dalibard2011colloquium},
\begin{equation}
i \hbar \frac{\partial \phi_1(\boldsymbol{r})}{\partial t}=\left[\frac{(\boldsymbol{p}-\boldsymbol{A})^2}{2 M}+V(\boldsymbol{r})+\varepsilon_1 + W\right] \phi_1(\boldsymbol{r}) .
\end{equation}
Now the reduced vector potential is denoted by $\boldsymbol{A}=i\hbar\langle\chi_1|\nabla \chi_1\rangle$. This is associated with the Born-Oppenheimer approximation, where the external dynamics are separated from the internal state. Here, the additional scalar potential $W(\boldsymbol{r})={\hbar^2}\sum_{m=2}^N|\langle\chi_m|\chi_1\rangle|^2/{2M}$ appears due to the nonadiabatic transitions from other dressed states, which can be absorbed in the potential field $V(\boldsymbol{r})$. 

\begin{figure}[tbp]
\centering
\includegraphics[width=1.0\linewidth]{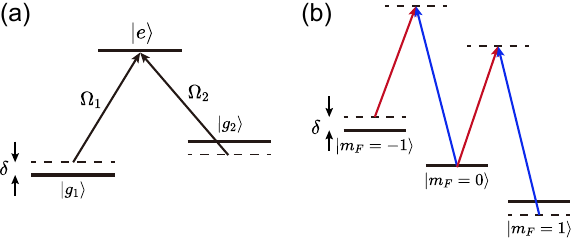}
\caption{(a) Level structure of $\Lambda$-type atom with a dark state determined by Rabi frequencies. (b) Raman dressing scheme for Zeeman sublevels of the $^{87}\mathrm{Rb}$ atom.}
\label{Fig10}
\end{figure}

In experiments, the $\Lambda$-type atomic-level structure is one of the most widely used configurations for generating light-induced gauge fields, in which the internal states of atoms adiabatically follow the dark state, as mentioned above. The dark state is formed by a linear combination of two ground states, and it suppresses the heating problem caused by spontaneous emission from the excited state. As shown in Fig.~\ref{Fig10}(a), the light-atom coupling term is now given by
\begin{equation}
\mathbb{M}=\hbar\delta(|g_2\rangle\langle g_2|-|g_1\rangle\langle g_1|)+\sum_{i=1}^2\hbar(\Omega_i|g_i\rangle\langle e|+\mathrm{H.c.}),
\end{equation} 
where the two laser beams induce coherent transitions between ground states $|g_i\rangle$ and the excited state $|e\rangle$, with position-dependent Rabi frequencies $\Omega_i(\boldsymbol{r})=|\Omega_i|e^{i\phi_i}$ ($i=1,2$). When the excited state $|e\rangle$ is in resonance, i.e., $\delta=0$, the system manifests a dark state $|D\rangle=(\Omega_2|g_1\rangle-\Omega_1|g_2\rangle)/\Omega$, and $\Omega=\sqrt{|\Omega_1|^2+|\Omega_2|^2}$. For atoms initially prepared in this dark state, the associated geometric gauge potential is given by~\cite{dalibard2011colloquium,Goldman_2014}
\begin{equation}
\begin{gathered}
\boldsymbol{A}=i\hbar\langle D(\boldsymbol{r})|\nabla  D(\boldsymbol{r})\rangle=-\hbar\frac{ |\zeta|^2 \nabla \phi}{1+|\zeta|^2},\\
\zeta=\Omega_1(\boldsymbol{r})/\Omega_2(\boldsymbol{r})=|\zeta|\mathrm{exp}(i\phi),
\end{gathered}
\end{equation}
where $|\zeta|=|\Omega_1|/|\Omega_2|$ and $\phi=\phi_1-\phi_2$ denote the amplitude ratio of two Rabi frequencies and the phase difference, respectively. In addition, spatial gradients of these two parameters must be nonzero to induce a nontrivial synthetic flux along a closed loop of atomic motion. This can be realized using two laser beams with orbital angular momentum~\cite{juzeliunas2004slow}. For bosonic ultracold atoms hopping between individual traps, the light-induced SGF can be engineered to realize a boson-SGF model with chiral flows. Alternatively, one can apply the Raman dressing scheme of a ladder-type atom where the excited state is replaced with another ground state, as sketched in Fig.~\ref{Fig10}(b). This configuration can be realized with alkali ultracold atoms, like the Raman-dressed $^{87}\mathrm{Rb}$ Bose-Einstein condense in the $F=1$ ground-state manifold~\cite{lin2009bose}. To create Zeeman sublevels $|m_F=0,\pm1\rangle$, the degeneracy is lifted by an external bias magnetic field. The spatially varying two-photon detuning $\delta$ in the Raman coupling generates a synthetic vector potential for the atoms, leading to position-dependent eigenstates. This Raman dressing scheme can also be applied to a two-level system; however, intrinsic limitations arise concerning the spatial range of SGFs and the balance between off-resonant scattering and Raman coupling~\cite{spielman2009raman}.

% Both the use of geometric phases and rotating the system can provide artificial magnetism.

% SGF is obtained in noninertial frame of reference containing both rotation and shaking optical lattices. Except the modulation of Hamiltonian in a noninertial frame of reference, SGF also emerges in an atomic Hamiltonian with a geometric gauge potential in light-matter coupling systems. As neutral atoms adiabatically evolve at spatial-dependent dressed states, the geometric phase acquired by neutral atoms implies an effective magnetic field acting on these neutral particles. Moreover, this scheme can be also extended to generate non-Abelian gauge fields.

\section{Conclusion and Outlook}
\label{Sec:Con}

In conclusion, we investigate the boson-SGF model and show the chiral excitation flow in the three-node case with a broken TR symmetry; however, the pure SGF model cannot drive the chiral dynamics when it extends to the four-node case. By introducing the auxiliary node to couple to network nodes, we recreate the chiral flow in the four-node boson-aSGF network. The excitation dynamics of this four-node boson-aSGF network suggest that it can exhibit chiral flow only when the synthetic flux of each triangular substructure breaks TR symmetry, i.e., $\theta_\vartriangle\neq p\pi$ ($p\in\mathbb{Z}$). This links the well-studied three-node chiral flow and the boson-aSGF model for creating perfect chiral flows in multinode networks, where these triangular substructures are also retained, with $(p+1/2)\pi$ ($p\in\mathbb{Z}$) synthetic flux. When we generalize this framework to the spin or two-level atomic system where TR symmetry is preserved, it demonstrates opposite chiral dynamics for the spin-up and spin-down states. Further, we introduce a universal scheme for constructing $n$-node chiral-flow Hamiltonians with the aSGF model and provide illustrative examples for odd and even $n$ values. During this process, we find the criteria for a Hamiltonian to support the chiral flow in a specific excitation subspace, which indicates the requirement for a symmetric and equally spaced energy spectrum and a complete set of chiral modes. Our framework for finding chiral flows applies to various systems such as bosons, spins, and two-level atoms. It is also applicable to different types of interactions, including the NN complex hopping forming a two-body Hamiltonian, as well as the three-body interactions like SCI and ASI in spin systems. Nonetheless, to realize the chiral flow in a larger network, the number of required hoppings scales as $n^2/2$, showing significant complexity growth. Therefore, we propose a ladder network with the complexity scaling as $n$. This can serve as a more practical scheme for unidirectional transmission between remote emitters. In addition, the imperfections of networks are also discussed, where the fidelity of chiral flow shows robustness in the presence of hopping disorder but is sensitive to the inhomogeneity of node (boson mode or atomic) frequency.

The multinode aSGF model and the extension of the four-node network can be realized with superconducting circuits by employing Floquet modulation, offering a practical approach to implementing perfect state transfer in circuit QED systems. The magnetic photonic lattice can also mimic this model to a certain extent with atomic emitters. Moreover, synthetic gauge fields in cold-atom systems provide a feasible route for implementing both boson- and spin-aSGF models. The experimental feasibility of our scheme, along with its tolerance to imperfections, makes it a practical and promising approach for realizing chiral state transfer and unidirectional entanglement distribution. We also expect that our proposal could be used to develop nonreciprocal quantum devices~\cite{SciPostPhys.Lect.Notes2022Clerk}, such as $n$-node circulators, and benefit future quantum technologies.
% ---------------------------------------------------------------------------
% ---------------------------------------------------------------------------

\section{Acknowledgments}

This work is supported by the National Natural Science Foundation of China (Grant No. 12375025 and No. 11874432) and the National Key R\&D Program of China (Grant No. 2019YFA0308200).
% ---------------------------------------------------------------------------
%----------------------------------------------------------------------------

\appendix

\section{The analytical solution of four-node boson-SGF model with arbitrary gauge phase}
\label{asec:4nodeSGF_nocentral}

The single-excitation dynamics of the four-node boson-SGF model under a symmetric gauge can be analytically solved for arbitrary synthetic flux $\theta_4$. Here, we denote the nearest-neighboring hopping phase as $\theta=\theta_4/4$. The four-node boson-SGF Hamiltonian in the single-excitation subspace then reads
\begin{equation}
\mathcal{H}_{{\rm SGF}}^{4}/J_0=\left(\begin{matrix}0 & e^{i\theta} & 0 & e^{-i\theta}\\
e^{-i\theta} & 0 & e^{i\theta} & 0\\
0 & e^{-i\theta} & 0 & e^{i\theta}\\
e^{i\theta} & 0 & e^{-i\theta} & 0
\end{matrix}\right).
\end{equation}
Here, the single-excitation basis kets are $\{|j\rangle\equiv a^\dagger_j|G\rangle\}$, with $j=1, 2, 3, 4$. The eigenvalues $\lambda_k$ and corresponding eigenvectors $\mathbf{f}^k$ of $\mathcal{H}^4_\mathrm{SGF}$ are given by
\begin{equation}\label{equ:4node_noC_eigen}
\begin{cases}
\mathbf{f}^{1}=\frac{1}{2}\left(1,1,1,1\right)^T,\lambda_{1}=2\cos\theta,\\
\mathbf{f}^{2}=\frac{1}{2}\left(-1,1,-1,1\right)^T,\lambda_{2}=-2\cos\theta,\\
\mathbf{f}^{3}=\frac{1}{2}\left(-i,-1,i,1\right)^T,\lambda_{3}=2\sin\theta,\\
\mathbf{f}^{4}=\frac{1}{2}\left(i,-1,-i,1\right)^T,\lambda_{4}=-2\sin\theta.
\end{cases}
\end{equation}
When $\theta=\pi/2$, $\mathbf{f}^1$ and $\mathbf{f}^2$ become the degenerate zero-energy states. 

By setting the initial state as $|\psi(0)\rangle=a_1^\dagger|G\rangle=(1,0,0,0)^T$, the system's dynamics are determined by the Schr\"odinger equation, with $|\psi(t)\rangle=e^{-iH^4_{\rm SGF}t}|\psi(0)\rangle$. We denote $|\psi(t)\rangle=\sum_jC^j_e(t)|j\rangle$, and the probability amplitudes can be written as
\begin{equation}\label{equ:SGFCje_gen}
C^{j}_e\left(t\right)=\sum_{k}\left(f_{1}^{k}\right)^{*}f_{j}^{k}e^{-iE_{k}t},E_k=J_0\lambda_k,
\end{equation}
where $f_j^k = \langle j|\mathbf{f}^k\rangle$. Substituting the eigenvalues and eigenvectors from Eq.~\eqref{equ:4node_noC_eigen} into the above equation, the population $P^j_e(t)\equiv |C^j_e(t)|^2$ can be obtained~\cite{PhysicaE2004Taut}:
\begin{equation}
\begin{split} & P_{e}^{1}(t)=\frac{1}{4}\left(\cos\left[2\cos(\theta)t\right]+\cos\left[2\sin(\theta)t\right]\right)^{2},\\
 & P_{e}^{2}(t)=P_{e}^{4}(t)=\frac{1}{4}\left(\sin^{2}\left[2\sin(\theta)t\right]+\sin^{2}\left[2\cos(\theta)t\right]\right),\\
 & P_{e}^{3}(t)=\frac{1}{4}\left(\cos\left[2\cos(\theta)t\right]-\cos\left[2\sin(\theta)t\right]\right)^{2}.
\end{split}
\end{equation}
The populations of the second and fourth nodes are always the same, which reveals the same dynamical process, as shown in Figs.~\ref{Fig1}(e) and \ref{Fig1}(f). In particular, when $\theta_4=\pi$ and $\theta=\pi/4$, the third node can be considered as a dark site whose population vanishes all the time, with $P^3_{e}(t, \theta_4=\pi)=0$, as shown in Fig.~\ref{Fig1}(e). Intuitively, the transition amplitude of the $1\rightarrow 3$ process can be written as $t_{13}\propto (\exp(i\pi/2)+\exp(-i\pi/2))=0$, with destructive interference from two paths. 

% ---------------------------------------------------------------------------
% ---------------------------------------------------------------------------

\section{The dynamics of \texorpdfstring{$n$}.-node networks}
\label{asec:5nodeand6node_SGF}

In this appendix, we study the dynamics of $n$-node networks and show the destruction of chiral flows for the networks whose number of nodes is greater than four, i.e., $n\geq 4$.

We consider the $n$-node boson-SGF model in which each node has a $\pi/2$ hopping phase from its next node, and the Hamiltonian reads
\begin{equation}
H_\mathrm{SGF}^n=J_0 \sum_{j=1}^n (i a_j^\dagger a_{j+1}+\mathrm{H.c.}),
\end{equation}
where the periodic boundary condition $a_{n+1}=a_1$ is used. In the single-excitation subspace with Hamiltonian $\mathcal{H}_\mathrm{SGF}^n$, the eigenstates are plane waves,
\begin{figure}[tbp]
\centering
\includegraphics[width=1.0\linewidth]{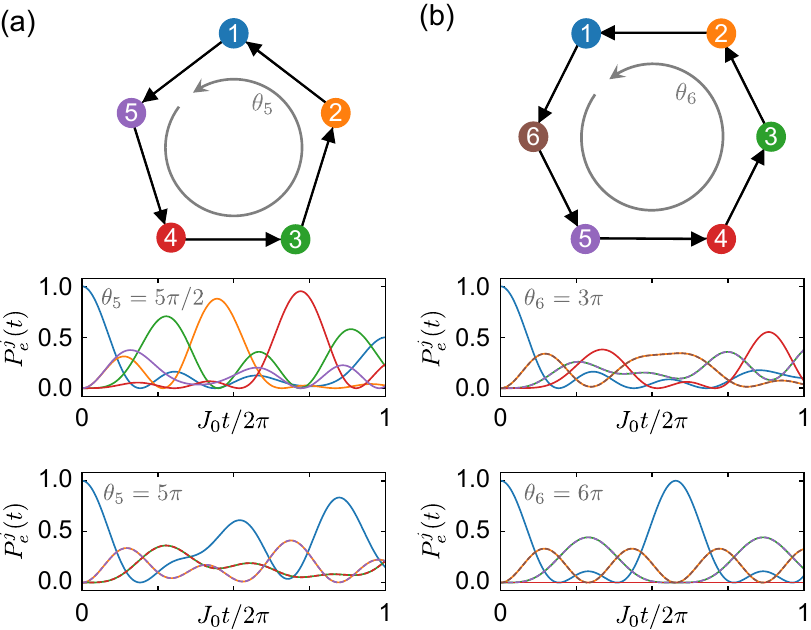}
\caption{Evolution of excited-state population $P_e^j(t)$ of nodes in (a) five-node and (b) six-node boson-SGF models. No chiral flow appears, regardless of whether the synthetic flux breaks TR symmetry.}
\label{FigB1}
\end{figure}
\begin{equation}\label{equ:SGFeigenstate_gen}
|\varphi_{m}\rangle=\sum_{j=1}^n e^{i j k_m}a_j^\dagger|G\rangle/\sqrt{n},\quad m\in\mathbb{Z},-\frac{n}{2}<m\leq \frac{n}{2},
\end{equation}
with an energy spectrum given by $E_m=-2J_0\sin k_m$, where $k_m=2\pi m/n$ are the wave numbers within the first Brillouin zone (BZ).  For instance, when $n=4$, the wave numbers with $m=0,2,-1$ and $1$ reproduce the four eigenstates in Eq.~\eqref{equ:4node_noC_eigen}. We refer to these eigenstates as chiral modes since they have chiral phase shifts $2\pi m/n$ between nearest sites. Note that the system has an internal chiral symmetry $\mathcal{C}$, with $\mathcal{C}^{-1}\mathcal{H}_\mathrm{SGF}^n \mathcal{C}=-\mathcal{H}_\mathrm{SGF}^n$, resulting in a symmetric energy spectrum~\cite{Asboth2016ShortCourseTopological}. The chiral operators for even and odd numbers of nodes are given by
\begin{equation}
\mathcal{C} = \bigoplus_{j=1}^{n/2} \sigma_z^j \quad\text{and} \quad\mathcal{C}=
\begin{pmatrix}
0 & 0&\cdots& 0 &1\\
0 & 0 &\cdots&1&0 \\
\vdots&\vdots&\reflectbox{$\ddots$} &\vdots&\vdots\\
0 &1&\cdots&0 &0 \\
1& 0&\cdots&0 &0
\end{pmatrix},
\end{equation}
respectively.

\begin{figure}[tbp]
\centering
\includegraphics[width=0.9\linewidth]{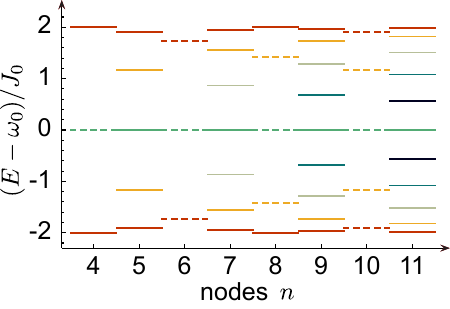}
\caption{Energy spectrum of $n$-node boson-SGF model for $n$ values from 4 to 11. The dashed lines indicate degenerate energies with two degenerate eigenstates.}
\label{FigB2}
\end{figure}

In Figs.~\ref{FigB1}(a) and \ref{FigB1}(b), we plot the dynamics of five-node and six-node boson-SGF models, finding that the perfect chiral flow cannot be supported within these networks, even when the hopping phase is $\pi/2$. To understand the underlying reasons for this, we categorize the boson-SGF networks into odd and even classes based on the value of $n$. When $n$ is odd, the presence of chiral symmetry gives rise to a single zero-energy state, $|\varphi_{0}\rangle=\sum_j^n a_j^\dagger|G\rangle/\sqrt{n},~E_0=0$, corresponding to $k_m=0$. Here, we assume that the excitation is initially on the first node. Then, using Eq.~\eqref{equ:SGFCje_gen}, the evolution of excitation on $j$th node $P_e^j(t)=|C_e^j(t)|^2$ can be obtained, with the probability amplitude given by
\begin{equation}\label{oddnet}
C_e^j(t)=\sum_{k_m\in \mathrm{BZ}} \frac{2}{n}\cos\left[E_m t-k_m(j-1)\right]+\frac{1}{n},
\end{equation}
which implies that realizing the perfect chiral flow requires a harmonic energy spectrum, i.e., $E_m=m E_1$, while this condition cannot be satisfied when $n>4$, as shown in Fig.~\ref{FigB2}. For the even class, there are two degenerate zero-energy states, denoted as 
\begin{equation}
|\varphi_{0,+}\rangle=\sqrt{\frac{2}{n}}\sum_{j=1}^{n/2}a_{2j-1}^\dagger|G\rangle,\; |\varphi_{0,-}\rangle=\sqrt{\frac{2}{n}}\sum_{j=1}^{n/2}a_{2j}^\dagger|G\rangle,
\end{equation}
which are the linear combinations of the two zero-energy eigenstates in Eq.~\eqref{equ:SGFeigenstate_gen}, i.e., $|\varphi_{0,\pm}\rangle=(|\varphi_0\rangle\pm|\varphi_{n/2}\rangle)/\sqrt{2}$. Therefore, we can deduce the probability amplitude
\begin{equation}
C_{e}^{j}(t)=\sum_{k_m\in\mathrm{BZ}} \frac{2}{n}\cos\left[E_{m}t-k_m(j-1)\right]+\frac{2}{n}\delta_{j,{\rm odd}},
\end{equation}
where $\delta_{j,{\rm odd}}=1$ if $j$ is odd and $\delta_{j,{\rm odd}}=0$ if $j$ is even. In this sense, the wave function of zero-energy eigenstate $|\varphi_{0,-}\rangle$ has no distribution on all odd sites, including the first site. This absence of amplitude leads to imperfect oscillations at all even sites, with maximum populations $\max[P_e^{j=2\mathbb{N}}(t)]\leq (1-2/n)^2$. Furthermore, the energy spectrum in Fig.~\ref{FigB2} also shows degeneracies when $n$ is even, which makes it impossible to fulfill the harmonicity condition $E_m=m E_1$.

% ---------------------------------------------------------------------------
% ---------------------------------------------------------------------------
% ---------------------------------------------------------------------------

\section{Analytical solutions of four-node and six-node boson-aSGF models}
\label{asec:deri}

Since the four-node and six-node boson-aSGF models have simple explicit expressions of eigenstates and energy levels, here, we derive the analytic solutions for the dynamics of the node excited-state populations $P_e^j(t)$. In the single-excitation subspace, the Hamiltonian~\eqref{equ:SGFCmodel} can be written in the following matrix form with $n=4$:
\begin{equation}\label{equ:n_node_Hcal_aSGF}
    \begin{aligned}
        &\mathcal{H}_\mathrm{aSGF}^{n}/J_{0}=\\
        &\begin{pmatrix}0 & e^{i\theta_{12}} & 0 & \cdots & 0 & e^{-i\theta_{n1}} & \beta_{c}\\
            e^{-i\theta_{12}} & 0 & e^{i\theta_{23}} & \cdots & 0 & 0 & \beta_{c}\\
            0 & e^{-i\theta_{23}} & 0 & \cdots & 0 & 0 & \beta_{c}\\
            \vdots & \vdots & \vdots & \ddots & \vdots & \vdots & \vdots\\
            0 & 0 & 0 & \cdots & 0 & e^{i\theta_{n-1,n}} & \beta_{c}\\
            e^{i\theta_{n1}} & 0 & 0 & \cdots & e^{-i\theta_{n-1,n}} & 0 & \beta_{c}\\
            \beta_{c} & \beta_{c} & \beta_{c} & \cdots & \beta_{c} & \beta_{c} & 0
            \end{pmatrix},
    \end{aligned}
\end{equation}
where the basis kets are $\{|j\rangle\equiv a^\dagger_j|G\rangle, |c\rangle=c^\dagger|G\rangle\}$ with $j=1,\dots,n$. 

To solve the four-node boson-aSGF model with hopping phase $\theta_{j,j+1}=\pi/2$, we change the basis kets from $\{|1\rangle,|2\rangle,|3\rangle,|4\rangle\}$ to $\{|\alpha\rangle,|\beta\rangle,|\gamma\rangle,|\delta\rangle\}$, where the latter set consists of the eigenstates of the four-node SGF Hamiltonian. The unitary transformation is given by
\begin{equation}
\begin{pmatrix}
\vspace{0.5ex} a_\alpha^{\dagger} \\
\vspace{0.5ex} a_\beta^{\dagger} \\
\vspace{0.5ex} a_\gamma^{\dagger} \\
\vspace{0.5ex} a_\delta^{\dagger}
\end{pmatrix}=\begin{pmatrix}
\vspace{0.5ex} \frac{i}{2} & -\frac{1}{2} & -\frac{i}{2} & \frac{1}{2} \\
\vspace{0.5ex} -\frac{i}{2} & -\frac{1}{2} & \frac{i}{2} & \frac{1}{2} \\
\vspace{0.5ex} 0 & \frac{1}{\sqrt{2}} & 0 & \frac{1}{\sqrt{2}} \\
\vspace{0.5ex} \frac{1}{\sqrt{2}} & 0 & \frac{1}{\sqrt{2}} & 0
\end{pmatrix}\begin{pmatrix}
\vspace{0.5ex} a_1^{\dagger} \\
\vspace{0.5ex} a_2^{\dagger} \\
\vspace{0.5ex} a_3^{\dagger} \\
\vspace{0.5ex} a_4^{\dagger}
\end{pmatrix},
\end{equation}
and the Hamiltonian in the new basis is block-diagonal,
\begin{equation}
    \widetilde{\mathcal{H}}_\mathrm{aSGF}^{4}/J_{0}=\begin{pmatrix}-2\\
        & 2\\
        &  & 0 &  & \sqrt{2}\beta_c\\
        &  &  & 0 & \sqrt{2}\beta_c\\
        &  & \sqrt{2}\beta_c & \sqrt{2}\beta_c & 0
       \end{pmatrix}.
\end{equation}
Hence, the dynamics can be separated into two independent subspaces with the following Schr\"odinger equations and initial conditions: 
\begin{gather}
    i\partial_{t}\begin{pmatrix}C_{\alpha}\\
        C_{\beta}
        \end{pmatrix}=J_{0}\begin{pmatrix}-2\\
         & 2
        \end{pmatrix}\begin{pmatrix}C_{\alpha}\\
        C_{\beta}
        \end{pmatrix},\left\{\begin{array}{l}{C_{\alpha}(0)=-i/2} \\{C_{\beta}(0)=i/2}\end{array}\right.,\\
    i\partial_{t}\begin{pmatrix}C_{\gamma}\\
        C_{\delta}\\
        C_{c}
        \end{pmatrix}=J_{0}\sqrt{2}\beta_c\begin{pmatrix}0 &  & 1\\
            & 0 & 1\\
        1 & 1 & 0
        \end{pmatrix}\begin{pmatrix}C_{\gamma}\\
        C_{\delta}\\
        C_{c}
        \end{pmatrix},C_{\delta}\left(0\right)=\frac{1}{\sqrt{2}},
\end{gather}
where the explicit form of the state ket is written as $|\psi\left(t\right)\rangle=C_{\alpha}(t)|\alpha\rangle+C_{\beta}(t)|\beta\rangle+C_{\gamma}(t)|\gamma\rangle+C_{\delta}(t)|\delta\rangle+C_{c}(t)|c\rangle$. In the decoupled $\alpha\beta$ subspace, the Hamiltonian is diagonal with energy $\pm 2J_0$, leading to the free evolution of states $|\alpha\rangle$ and $|\beta\rangle$; however, in the $\gamma\delta c$ subspace, the zero-energy eigenstates of the four-node boson-SGF Hamiltonian $|\gamma\rangle$ and $|\delta\rangle$ interact with the auxiliary mode $|c\rangle$. After solving these time-dependent coefficients, we return to basis $\{|1\rangle,|2\rangle,|3\rangle,|4\rangle,|c\rangle \}$ and obtain the probability amplitudes for each node, 
\begin{equation}
    \begin{aligned}
        C_e^{1}\left(t\right)&=\frac{1}{4}\left[1+\cos\left(2J_{0}\beta_ct\right)+2\cos\left(2J_{0}t\right)\right],\\
        C_e^{2}\left(t\right)&=-\frac{1}{4}\left[1-\cos\left(2J_{0}\beta_ct\right)+2\sin\left(2J_{0}t\right)\right],\\
        C_e^{3}\left(t\right)&=\frac{1}{4}\left[1+\cos\left(2J_{0}\beta_ct\right)-2\cos\left(2J_{0}t\right)\right],\\
        C_e^{4}\left(t\right)&=-\frac{1}{4}\left[1-\cos\left(2J_{0}\beta_ct\right)-2\sin\left(2J_{0}t\right)\right].
    \end{aligned}
\end{equation}
More explicitly, they can be written in a compact form, as given in Eq. \eqref{equ:4nodespopu}. 

Similarly, in the six-node case with $\theta_{j,j+1}=\pi/2$, we use the unitary matrix
\begin{equation}\label{equ:6plus1model_Tran}
    U=\left(
\begin{array}{cccccc}
\vspace{0.5ex} \frac{i}{2} & -\frac{1}{\sqrt{3}} & -\frac{i}{2} & \frac{1}{2 \sqrt{3}} & 0 & \frac{1}{2 \sqrt{3}} \\
\vspace{0.5ex} -\frac{1}{2 \sqrt{3}} & 0 & -\frac{1}{2 \sqrt{3}} & -\frac{i}{2} & \frac{1}{\sqrt{3}} & \frac{i}{2} \\
\vspace{0.5ex} -\frac{i}{2} & -\frac{1}{\sqrt{3}} & \frac{i}{2} & \frac{1}{2 \sqrt{3}} & 0 & \frac{1}{2 \sqrt{3}} \\
\vspace{0.5ex} -\frac{1}{2 \sqrt{3}} & 0 & -\frac{1}{2 \sqrt{3}} & \frac{i}{2} & \frac{1}{\sqrt{3}} & -\frac{i}{2} \\
\vspace{0.5ex} 0 & \frac{1}{\sqrt{3}} & 0 & \frac{1}{\sqrt{3}} & 0 & \frac{1}{\sqrt{3}} \\
\vspace{0.5ex} \frac{1}{\sqrt{3}} & 0 & \frac{1}{\sqrt{3}} & 0 & \frac{1}{\sqrt{3}} & 0 \\
\end{array}
\right)
\end{equation}
to block-diagonalize $\mathcal{H}_\mathrm{aSGF}^6$, i.e., change the basis from $\{|1\rangle,\dots,|6\rangle,|c\rangle\}$ to $\{|\alpha\rangle,|\beta\rangle,|\gamma\rangle,|\delta\rangle,|\varepsilon\rangle,|\zeta\rangle, |c\rangle \}$, which are the eigenstates of six-node SGF Hamiltonian $\mathcal{H}_\mathrm{SGF}^6$. After the transformation, we obtain the following form:
\begin{equation}
    \widetilde{\mathcal{H}}_\mathrm{aSGF}^{6}/J_{0}=\begin{pmatrix}-\sqrt{3} &  &  &  &  &  & 0\\
        & -\sqrt{3} &  &  &  &  & 0\\
        &  & \sqrt{3} &  &  &  & 0\\
        &  &  & \sqrt{3} &  &  & 0\\
        &  &  &  & 0 &  & \sqrt{3}\beta_{c}\\
        &  &  &  &  & 0 & \sqrt{3}\beta_{c}\\
       0 & 0 & 0 & 0 & \sqrt{3}\beta_{c} & \sqrt{3}\beta_{c} & 0
       \end{pmatrix}.
\end{equation}
The dynamics can be likewise separated into two parts. The first part corresponds to the $\alpha\beta\gamma\delta$ subspace in the upper-left $4\times 4$ block, with a diagonal matrix that indicates the simple free evolution. The second part is restricted to the $\varepsilon\zeta c$ subspace in the lower-right $3\times 3$ block, which denotes the coupling between zero-energy states and the auxiliary mode. After solving the coefficients of the time-dependent state $|\psi\left(t\right)\rangle=C_{\alpha}(t)|\alpha\rangle+C_{\beta}(t)|\beta\rangle+C_{\gamma}(t)|\gamma\rangle+C_{\delta}(t)|\delta\rangle+C_\varepsilon(t)|\varepsilon\rangle+C_\zeta(t)|\zeta\rangle+C_{c}(t)|c\rangle$, and transforming back to the original basis $\{|1\rangle,\dots,|6\rangle,|c\rangle\}$, we obtain 
\begin{equation}
\label{equ:6plus1model_analytic}
\begin{split}
    &P^1_e(t)=\frac{1}{36}\left[1+\cos(\sqrt{6}J_0\beta_ct)+4\cos(\sqrt{3}J_0t)\right]^2,\\
    &P^2_e(t)=\frac{1}{36}\left[1-\cos(\sqrt{6}J_0\beta_ct)+2\sqrt{3}\sin(\sqrt{3}J_0t)\right]^2,\\
    &P^3_e(t)=\frac{1}{36}\left[1+\cos(\sqrt{6}J_0\beta_ct)-2\cos(\sqrt{3}J_0t)\right]^2,\\
    &P^4_e(t)=\frac{1}{36}\left[1-\cos(\sqrt{6}J_0\beta_ct)\right]^2,\\
    &P^5_e(t)=\frac{1}{36}\left[1+\cos(\sqrt{6}J_0\beta_ct)-2\cos(\sqrt{3}J_0t)\right]^2,\\
    &P^6_e(t)=\frac{1}{36}\left[1-\cos(\sqrt{6}J_0\beta_ct)-2\sqrt{3}\sin(\sqrt{3}J_0t)\right]^2.
\end{split}
\end{equation}
This result suggests that the introduction of an auxiliary node is unable to create a perfect chiral flow in the six-node network, regardless of the coupling strength $\beta_c J_0$.
% ---------------------------------------------------------------------------
%----------------------------------------------------------------------------

\section{Dynamics of the extended four-node network}
\label{ExtendedNetwork}

In the extended four-node network, the energy spectrum is not equally spaced, which induces accumulated errors as the number of network nodes increases. To evaluate the imperfection of the chiral flow and investigate its origin, here we adopt a nonperturbative analysis of the system's propagator,
\begin{equation}
K_+(t,t')=U(t,t')\theta(\tau),\quad\tau=t-t',
\end{equation}
where $U(t,t')$ is the time-evolution operator and $\theta(\tau)$ denotes the Heaviside function. The propagator is related to the resolvent of the Hamiltonian $H_\mathrm{ex}^{N}$ by the Fourier transform
\begin{equation}
K_{+}(\tau)=-\frac{1}{2 \pi i} \int_{-\infty}^{+\infty} \mathrm{d} \omega \mathrm{e}^{-i \omega \tau / \hbar} G_{+}(\omega),
\end{equation}
where the resolvent is given by
\begin{equation}
G_+(\omega)\equiv\lim_{\eta\to0_+}G(\omega+i\eta)=\lim_{\eta\to0_+}\frac1{\omega-H_\mathrm{ex}^{N}+i\eta}.
\end{equation}

\begin{figure*}[tbp]
\centering
\includegraphics[width=1.0\linewidth]{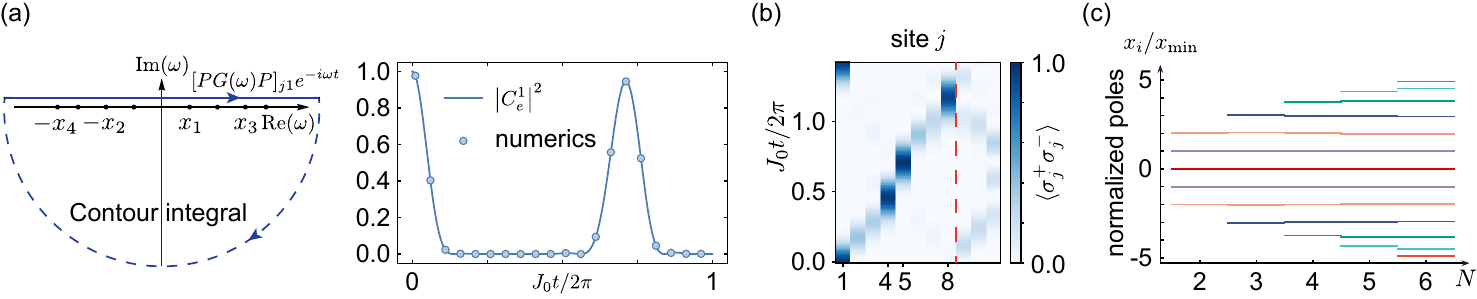}
\caption{(a) Schematic of the contour integral and a plot showing the population of the first node. Here, the solid line denotes the analytical expression for $P_e^1$, and the dots are numerical results obtained by solving the Schrödinger equation. (b) Time evolution of the excitation on each site for $N=4$. The region to the right of the red dashed line represents the evolution of auxiliary nodes. (c)~Normalized poles (with an $\omega_0$ shift) for different $N$ values. The unequally spaced poles result in the imperfection of fidelity.}
\label{Fig6}
\end{figure*}

To proceed, we divide the network-hopping Hamiltonian into two parts, $H_\mathrm{ex}^{N}=H_0+V_I$, as depicted by the solid and dashed lines in Fig.~\ref{Fig5}(a). The unperturbed Hamiltonian $H_0$ contains the free Hamiltonian of all nodes as well as the hoppings between the non-corner nodes, and the interaction Hamiltonian $V_I$ describes the hoppings from the four corner nodes labeled as $\mathscr{C}_1$, $\mathscr{C}_2$, $\mathscr{C}_3$, and $\mathscr{C}_4$, as shown in Fig.~\ref{Fig5}(a). We constrain the dynamics within the subspace $\mathscr{E}_0$ formed by the excited states of corner nodes and define the projector as $P=\sum_{j=1}^4|\mathscr{C}_{j}\rangle\langle \mathscr{C}_{j}|$. Similarly, the projector of the complementary subspace is given by $Q=1-P$. Thus, the projection of $G(\omega)$ in subspace $\mathscr{E}_0$ is written as
\begin{equation}
P G(\omega) P=\frac{P}{\omega-P H_0 P-P \Sigma(\omega) P},
\end{equation}
where $\Sigma(\omega)$ denotes the self-energy and is given by
\begin{equation}
\Sigma(\omega)=V_I+V_I \frac{Q}{\omega-Q H_0 Q-Q V Q} V_I.
\end{equation}
We can verify that in our system, the self-energy only extends to its first order, i.e.,
\begin{equation}
\Sigma(\omega)=V_I+V_I \frac{Q}{\omega-H_0} V_I.
\end{equation}
\begin{table}[bp]
\caption{Optimal coupling parameters}
    \centering
    \begin{ruledtabular}
    \begin{tabular}{ccc}
        number of copies & coupling parameters & fidelity \\
    \midrule
        N=2 & $2$ & $\mathcal{F}=0.986$\\
        N=3 & $(2,2.26)$ & $\mathcal{F}=0.960$\\
        N=4 & $(2,2.33)$ & $\mathcal{F}=0.933$\\
        N=5 & $(2,2.41,2.43)$ & $\mathcal{F}=0.909$\\
        N=6 & $(2,2.47,2.52)$ & $\mathcal{F}=0.886$\\
        N=7 & $(2,2.51,2.59,2.67)$ & $\mathcal{F}=0.865$\\
        N=8 & $(2,2.54,2.62,2.72)$ & $\mathcal{F}=0.933$\\
        N=9 & $(2,2.58,2.69,2.86,2.88)$ & $\mathcal{F}=0.847$\\
        N=10 & $(2,2.61,2.72,2.92,2.96)$ & $\mathcal{F}=0.831$\\
    \end{tabular}
    \label{tab:OCP}
    \end{ruledtabular}
\end{table}
As the system is initially prepared at $|\psi(t=0)\rangle=|c_1\rangle$, the probability amplitudes of the edge nodes are given by
\begin{equation}
\begin{aligned}
C_e^{cj}(t)=&K_+(t)|\psi(0)\rangle\\
=&\sum\mathrm{Res}\left\{[PG(\omega)P]_{j1} e^{-i\omega t}\right\},
\end{aligned}
\end{equation}
with $j=1,2,3,4$. Because the system preserves the chiral symmetry, the real poles of $PG(\omega)P$ are distributed symmetrically. To elucidate the retrieval of excitation on the first node and identify the cause of the fidelity imperfection, here we study the case $N=4$. We find that the simple poles are given by $x_{i,\pm}=\omega_0\pm k\sqrt{2}J_0$ with $i=1,2,3$ and $x_{4,\pm}=\omega_0\pm 2\sqrt{7}J_0$. By substituting the poles and calculating the sum of residues, we can then obtain the excited-state population of each node $|C_e^{cj}|^2$. The population of the first node is given by
\begin{equation}\label{Pe1network}
\begin{aligned}
P_e^1=&\left[\frac{11}{56}+ \frac{3}{8} \cos(\sqrt{2} J_0 t)+\frac{11}{40} \cos(2 \sqrt{2} J_0 t)\right.  \\
&\qquad \left.+\frac{1}{8} \cos(3 \sqrt{2} J_0 t)+\frac{1}{35} \cos(2 \sqrt{7} J_0 t)\right]^2.
\end{aligned}
\end{equation}
The poles of the contour integral are plotted in Fig.~\ref{Fig6}(a), and it can be seen that the numerical result (dots) is consistent with our analytical expression (solid line). We find that each pair of poles $\pm x_i$ contributes to a cosine oscillation on the probability as in Eq.~\eqref{Pe1network}, and the imperfection of fidelity comes from the fact that these poles are nonuniformly distributed; i.e., the closed propagation path from $\mathscr{C}_1$ to itself contains interference with nonharmonic frequencies. Figure~\ref{Fig6}(b) shows that the leakage mainly falls into the auxiliary nodes. We emphasize that these results are independent of the relative value of $J_0/\omega_0$ and there is no approximation made here. This implies that these hopping rates can also enter the nonperturbative regime. For larger networks, the error will accumulate as $n$ increases, along with the decreasing fidelity. This can be seen from the normalized poles $x_i/|x_\mathrm{min}|$ in Fig.~\ref{Fig6}(c), where $x_\mathrm{min}=\mathrm{min}\{|x_i|,|x_i|\neq 0\}$ induces the lowest oscillation frequency of $P_e^j(t)$.
% ---------------------------------------------------------------------------
% ---------------------------------------------------------------------------

\section{Optimized ladder network}\label{Appendix:OLN}
In the unmodified ladder-network model, the fidelity of chiral flow decreases rapidly as the number of copies $N$ increases. This can be unfavorable for remote transmission between corner nodes, particularly in large networks. To improve the fidelity of chiral transmission, we transform this ladder network into a polygonal network and replace the identical coupling parameter $\beta_c$ between auxiliary nodes and network nodes with a set of coupling parameters $\{\beta_j|j=0,\cdots,\lceil N/2\rceil -1\}$. As shown in Fig.~\ref{Fig5}(d), this configuration maintains the identical treatment of corner nodes. After fixing the coupling parameters at both ends to $\beta_0=2$, we iterate through $\{\beta_j|j>0\}$ within a specific range to find the optimal parameters $\{\beta_0^m=2,\beta_1^m,\cdots,\beta_{\lceil N/2\rceil -1}^m\}$ for each $N$, as listed in Table~\ref{tab:OCP}. Intuitively, the coupling strength should increase with $j$, as the auxiliary nodes that are distant from corner nodes require stronger coupling strength to carry out their role. The optimized fidelity is shown in Fig.~\ref{Fig5}(e), which demonstrates a significant improvement.
\begin{figure}[htbp]
\centering
\includegraphics[width=0.93\linewidth]{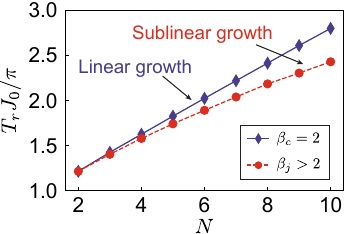}
    \caption{Scaling of the recovery time with and without the optimal parameters $\{\beta_j^m\}$. In both cases, there is a linear dependence on network size. Here, the blue solid line represents the fidelity from identical coupling, while the red dashed line denotes the results from optimal parameters.}
\label{FigB3}
\end{figure} 

The optimized ladder network also exhibits a shorter transmission time than the unmodified model. The circulation period of an excitation can be described by the recovery time $T_r$, which also characterizes the timescale for transferring an excitation between remote corner nodes. Although the recovery time shows a linear dependence on $N$, it does not increase in multiples (i.e., for a ladder network consisting of $N$ copies of the four-node aSGF model, $T_r<N\pi/J_0$). Instead, it increases much more slowly as the ladder network expands. The recovery time of the four-node aSGF model is $T_r=\pi/J_0$, while for a ladder network with $N=10$, the corresponding recovery time is less than $3\pi/J_0$, as illustrated in Fig.~\ref{FigB3}. When we apply the optimal parameters, there is a further reduction in $T_r$. As shown by the red dashed line in Fig.~\ref{FigB3}, $T_r$ scales sublinearly with $N$, which makes it promising for implementing remote transmission.

%%%%%%%

\bibliography{reference_MNCF.bib}

\end{document}